\documentclass{wstmp}
\usepackage[super,compress]{cite}
\usepackage{graphicx}
\usepackage{diagbox,xfrac}
\usepackage{color}

\newcommand{\imu}{{\rm i}}
\newcommand{\fract}[2]{{\textstyle\frac{#1}{#2}}}
\newcommand{\Vek}[1]{{\boldsymbol#1}}
\newcommand{\vek}[1]{\mbox{\footnotesize\boldmath$#1$\unboldmath}}
\newcommand{\dslash}{\partial \hskip -0.6em /}
\newcommand{\Dslash}{D \hskip -0.7em /}
\newcommand{\zr}[1]{\mbox{\hspace*{#1em}}}
\newcommand{\ID}{\mbox{{\sf 1}\zr{-0.16}\rule{0.04em}{1.55ex}\zr{0.1}}}

\begin{document}
\markboth{N. Graham, H. Weigel}{Quantum corrections to soliton energies}

\title{Quantum corrections to soliton energies}

\author{N. Graham}

\address{Department of Physics, Middlebury College\\
Middlebury, VT 05753, USA\\
ngraham@middlebury.edu}

\author{H. Weigel}

\address{Institute for Theoretical Physics, Physics Department\\
Stellenbosch University, Matieland 7602, South Africa\\
weigel@sun.ac.za}

\maketitle

\begin{abstract}
We review recent progress in the computation of leading quantum corrections
to the energies of classical solitons with topological structure, including 
multi-soliton models in one space dimension and string configurations in 
three space dimensions. Taking advantage of analytic continuation techniques
to efficiently organize the calculations, we show how quantum corrections
affect the stability of solitons in the Shifman-Voloshin model,
stabilize charged electroweak strings coupled to a heavy fermion
doublet, and bind Nielsen-Olesen vortices at the classical transition
between type I and type II superconductors.

\end{abstract}

%\ccode{PACS numbers:}

%\tableofcontents

\section{Introduction}

The existence of degenerate vacuum configurations in a classical field theory
allows for the possibility of topological soliton, or, more precisely,
solitary wave  solutions~\cite{Ra82}, which are solutions to the field
equations with localized energy densities. Because such solitons are
stabilized by global properties of the classical solution, one expects
them to remain stable when quantum effects are
included. However, in a variety of situations, quantum corrections can
alter the classical results in a significant way. These effects can be
formulated as the  shift in the zero-point (vacuum) energies of the
fluctuations when subjected to the potential induced by the
soliton. This vacuum polarization energy (VPE) is\footnote{Unless
  noted otherwise, we write formulas for boson fluctuations. Fermion
  fluctuations require an overall minus sign.}
\begin{equation}
\Delta E=\frac{1}{2}\sum_k\left(\omega_k-\omega_k^{(0)}\right)+E_{\rm CT}\,,
\label{eq:vpeI}
\end{equation}
where $\omega_k$ and $\omega_k^{(0)}$ are the energy eigenvalues of the 
fluctuations in the background of the soliton and the translationally invariant
vacuum, respectively, and $E_{\rm CT}$ is the counterterm
contribution, described below. The formal sum in Eq.~(\ref{eq:vpeI})
can be expressed as a discrete sum over bound states
plus a continuum integral over scattering states.
We note that Eq.~(\ref{eq:vpeI}) may also be taken as the starting point for 
studies of the Casimir force on conductors due to the exchange of
virtual photons \cite{MiltonBook,DalvitBook}, and so the VPE
is often also called the Casimir energy. For a charged soliton, one
can additionally include effects of occupied levels,  but such
contributions must be considered together with the VPE since they
appear at the same order in~$\hbar$.

Even when taking the difference of energy levels as in
Eq.~(\ref{eq:vpeI}), the sum over zero-point energies diverges, so the
theory must be regularized, and counterterm contributions $E_{\rm CT}$
must be added to render the VPE finite. In renormalizable theories,
the counterterms are  linear combinations of terms already present in the 
local Lagrangian before quantization. The coefficients of these terms are
determined from conditions on Green's functions that do not depend on
the field configuration under consideration.  These renormalization
conditions define masses and couplings in the quantum theory that are
fixed from experimental data.  Based on these inputs, one then obtains
unambiguous predictions from calculations in the renormalized theory.

Particularly because most applications in four spacetime dimensions
ultimately require numerical analysis, additional tools are required to pass 
from these formal expressions to tractable calculations with no cancellations 
of large or cutoff-dependent quantities. Without such tools, early
calculations  required extremely high
precision~\cite{Perry:1986fs}. As was already  recognized in the
earliest VPE calculations \cite{Dashen:1974ci,DHN}, scattering theory
methods can play an invaluable role in improving this  situation. In
this approach, the continuum part of the sum in Eq.~(\ref{eq:vpeI})
is rewritten as an integral over the change of the density of states, which 
in turn is related to the scattering phase shift. The VPE calculation is then
connected with standard renormalization procedures by identifying
contributions from the Born approximation of the phase shift
with the corresponding terms in the Feynman diagram expansion
for the VPE \cite{PhaseshiftsGeneral,Graham:2009zz}.

Even with these tools, however, the sums and integrals over the entire
spectrum of quantum fluctuations are still challenging numerically for 
phenomenologically relevant models.  As in other numerical computations 
in quantum field theory, it is helpful to use contour integration to 
shift the integral over the density of states to the imaginary momentum axis.  
One immediate benefit of this approach is that bound states no longer need 
to be identified explicitly, since their contributions are exactly
canceled by those from poles in the contour integral \cite{Bordag:1996fv}.
Furthermore, careful extension of the variable phase approach to
scattering theory \cite{variable} can enable one to replace
oscillating functions with decaying exponential functions (while
avoiding growing exponential functions) \cite{Graham:2002xq,Graham:2009zz},
significantly improving the efficiency of the numerical computation.

Owing to its fundamental importance in quantum field theory, there is
an extensive literature covering many approaches to and applications
of VPE calculations.  A necessarily incomplete summary includes
Green's function methods,
\cite{Li:1988hv,Moussallam:1989uk,Baacke:1989sb,Baacke:1991nh}
which allow for a similar Born approximation identification of the
divergent diagrams and counterterms; heat kernel techniques,
based on the proper-time representation of the determinant,
which can be used to provide long-wavelength approximations 
\cite{Ebert:1985kz,Avramidi:1990je,Avramidi:1995ik},
for sufficiently smooth background configurations, as well as exact results
\cite{Zuk:1990vx,Bordag:1994jz,Bordag:1996fv,Bordag:1998tg,Bordag:1999sf,Vassilevich:2003xt},
although the comparison to standard renormalization conditions is more
difficult in this approach; the Gel'fand-Yaglom method 
\cite{Gelfand:1959nq,ColemanBook,Parnachev:2000fz,Dunne:2006ct,Dunne:2008wh,Baacke:2008zx,Dunne:2012vv},
which obtains the energy sum in Eq.~(\ref{eq:vpeI}) from the solution
to a differential equation; the world-line formalism
\cite{Gies:2001zp,Gies:2001tj,Langfeld:2002vy,Gies:2003cv}, which 
uses more extensive numerical computation and as a result can
accommodate configurations without sufficient symmetry for a partial
wave expansion; and generalized derivative expansion methods, which
can yield both exact results when summed to all
orders\cite{Cangemi:1994by,Dunne:1999du,Chan:1996sa} and simpler
approximate results for slowly varying backgrounds
\cite{Aitchison:1984ys,Aitchison:1985pp,Bagger:1991pg,Bagger:1991ph}.

In this review, we begin in Sec.~\ref{sec:method} by recapitulating the
spectral methods approach for computing the VPE,
with emphasis on the effectiveness
of the imaginary momentum integration. We then discuss three situations 
where quantum corrections can qualitatively affect the properties of classical 
solitons with topological charge arising from a winding number. 
In Sec.~\ref{sec:Dim1}, we summarize how the VPE can destabilize multi-soliton 
solutions in one space dimension. In Sec.~\ref{sec:cstring}, we show how 
coupling to a heavy fermion can stabilize electroweak strings by allowing 
the energy of a fermions bound to the string to be lower than that of the
same number of free fermions. Finally, in Sec.~\ref{sec:vortices}
we discuss quantum corrections to Nielsen-Olesen vortices 
as a function of winding number, and show that in the BPS case of
equal gauge and Higgs masses, quantum corrections favor higher winding
over a corresponding number of isolated vortices with unit
winding, while in the classical model these energies are equal.

\section{Spectral methods}
\label{sec:method}

Spectral methods are the main tool to compute the VPE of static, extended
field configurations. These configurations induce a potential for small 
amplitude fluctuations, which are treated by standard techniques of
scattering  theory in quantum mechanics. They provide the bound state
energies, $\omega_j$,  which directly enter the VPE, as well as the
phase shifts $\delta(k)$ (or more 
generally the scattering matrix) as a function of the wave-number $k$ for single 
particle energies above threshold given by the mass $m$ of the fluctuating field. 
Those phase shifts parameterize the change in the density of continuum modes 
via the Friedel-Krein formula~\cite{Faulkner:1977aa}, 
\begin{equation}
\Delta\rho_\ell=\frac{1}{\pi}\frac{d\delta_\ell(k)}{dk}\,,
\label{eq:Krein}
\end{equation}
where we assume that the scattering potential has sufficient symmetry
to allow a partial wave expansion, represented by the $\ell$ index.

In turn, that change determines the continuum contribution to the VPE
\begin{align}
\Delta E&=\frac{1}{2}\sum_{j}^{\rm b.s.}\omega_j
+\frac{1}{2}\int_0^\infty dk \sum_\ell
\sqrt{k^2+m^2}\,\Delta\rho_\ell+E_{\rm CT}\cr
&=\frac{1}{2}\sum_{j}^{\rm b.s.}\omega_j
+\int_0^\infty \frac{dk}{2\pi}\sum_\ell
\sqrt{k^2+m^2}\,\frac{d\delta_\ell(k)}{dk}+E_{\rm CT}\,.
\label{eq:vpe1}
\end{align}
Here the partial wave sum includes any associated degeneracy 
factors, {\it e.g.} $2\ell+1$ for angular momentum in three space dimensions.

Eventually the analytic properties of scattering data allow for a more efficient 
computation of the VPE by introducing imaginary momenta $k=\imu t$ with $t\ge0$. 
Details of that approach have been reviewed elsewhere \cite{Graham:2009zz},
so here  we will focus on the main features for completeness.

\subsection{Scattering data}
\label{ssec:SD}

To compute the scattering data that enter Eq.~(\ref{eq:vpe1}), we
first write down the Schr\"odinger 
type equation for the radial part of
the fluctuation wavefunctions
\begin{equation}
\psi_{\ell,k}^{\prime\prime}(x)=-k^2\psi_{\ell,k}(x)
+\frac{1}{x^2}L^2\psi_{\ell,k}(x)+\sigma(x)\psi_{\ell,k}(x)\,,
\label{eq:SE1}
\end{equation}
where a prime indicates derivative with respect to the radial
coordinate $x$ and $L^2$ is square of the angular momentum eigenvalue. In one space dimension 
$x$ is position on the real axis with $L^2=0$, while in three
dimensions $x\ge0$ and
$L^2=\ell(\ell+1)$, for example. If the induced potential $\sigma(x)$ is attractive,
there are discrete bound state solutions with $\int dx |\psi_k(x)|^2\le\infty$ and
energy eigenvalues $\omega_j=\sqrt{m^2-\kappa_j^2}$ for $k^2=-\kappa_j^2<0$. In general
we consider a multi-channel problem, so that $\sigma(x)$ is matrix valued
while $\psi_{\ell,k}(x)$ is an array of $n$ wavefunctions. We can then define an
$n\times n$ matrix $\Psi_{\ell,k}(x)$, whose different columns represent independent
boundary conditions. In particular, introducing $\mathcal{H}_{\ell,k}(x)$ as the diagonal
matrix containing the free outgoing solutions, we parameterize
$\Psi_{\ell,k}(x)=\mathcal{F}_{\ell,k}(x)\cdot\mathcal{H}_{\ell,k}(x)$ and obtain the wave-equation
\begin{equation}
\mathcal{F}_{\ell,k}^{\prime\prime}(x)
=-2k\mathcal{F}_{\ell,k}(x)^\prime\cdot\mathcal{H}_{\ell,k}^\prime(x)\cdot\mathcal{H}_{\ell,k}^{-1}(x)
+\sigma(x)\cdot\mathcal{F}_{\ell,k}(x)\,.
\label{eq:SE2}
\end{equation}
Imposing the boundary condition $\lim_{x\to\infty}\mathcal{F}=\ID$ and observing that
the original wave-equation~(\ref{eq:SE1}) is real, we get the physical scattering solution
as the combination
\begin{equation}
\Psi^{\rm (sc)}_{\ell,k}(x)=\mathcal{F}_{\ell,k}^\ast(x)\cdot\mathcal{H}_{\ell,k}^\ast(x)
-\mathcal{F}_{\ell,k}(x)\cdot\mathcal{H}_{\ell,k}(x)\cdot\mathcal{S}(k)\,,
\label{eq:scat1}
\end{equation}
since $\mathcal{H}_{\ell,k}^\ast(x)$ asymptotically describes incoming waves.
The scattering matrix, $\mathcal{S}$, is extracted from the regularity 
condition\footnote{\label{ft:sym}The symmetric channel is one space dimension has 
$\lim_{x\to0}\Psi_{\ell,k}^{\prime{\rm(sc)}}(x)=0$ and a positive relative sign
in Eq.~(\ref{eq:scat1}).} $\lim_{x\to0}\Psi_{\ell,k}^{\rm(sc)}(x)=0$.
Finally the phase shift entering Eq.~(\ref{eq:vpe1}) is 
\begin{equation}
\delta_\ell(k)=\frac{1}{2\imu}\ln\left[{\rm det}S_\ell(k)\right]
=\frac{1}{2\imu}\ln \lim_{x\to0} 
\left[{\rm det}\left(\mathcal{F}_{\ell,k}^\ast(x)\mathcal{F}_{\ell,k}^{-1}(x)\right)\right]\,.
\label{eq:phase}
\end{equation}

\subsection{Renormalization}
So far our expression for the VPE has been quite formal since in Eq.~(\ref{eq:vpe1}) we 
still have to combine the infinities in the momentum integral and the
counterterms to obtain
a finite result. Conventionally the counterterm coefficients are determined from the
perturbative expansion of Green's functions via the computation of Feynman diagrams.
We can incorporate that scheme by first noting that there is a Feynman diagram expansion
for the quantum action in the presence of the potential $\sigma(x)$. The leading 
quantum correction is the sum of all one-loop diagrams
\begin{equation}
\mbox{Tr~ln}\left[\frac{D_2 + \sigma(x)}{D_2}\right]=
\parbox[b]{8.1cm}{
\raisebox{-0.88cm}{\includegraphics[width=8cm,height=2cm]{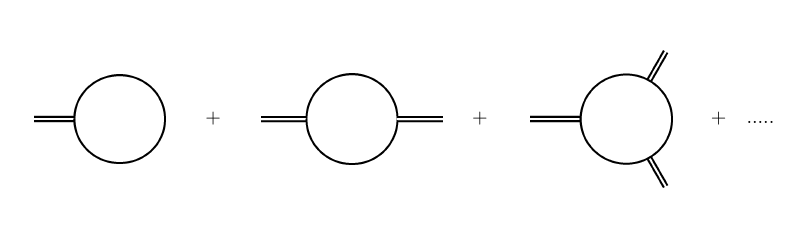}}} \,,
\label{eq:fdseries}
\end{equation}
where $D_2$ denotes the (covariant) second-order differential operator associated with the free 
wave-equation. The double lines represent insertions of the Fourier transform of 
$\sigma(x)$, so that the Feynman diagrams are given as integrals over those Fourier momenta and the loop momentum. 
Dividing by $D_2$ on the left-hand-side represents the subtraction of $\omega_k^{(0)}$ in 
Eq.~(\ref{eq:vpeI}). Since the potential is static, the energy is just the negative action
per unit time. Hence it is straightforward to associate an energy $E_{\rm FD}^{(n)}$ with the 
Feynman diagrams containing $n$ insertions of $\sigma(x)$.

On the other hand, we can expand the Jost solutions according to increasing orders in the 
potential by writing $\mathcal{F}=\ID+\mathcal{F}^{(1)}+\mathcal{F}^{(2)}+\ldots$ (omitting
labels for brevity). The $\mathcal{F}^{(n)}$ are subject to the differential equations \cite{variable}
\begin{equation}
\mathcal{F}^{(1)\prime\prime}=
-2\mathcal{F}^{(1)\prime}\cdot\mathcal{H}^\prime\cdot\mathcal{H}^{-1}+\sigma\,,
\quad
\mathcal{F}^{(2)\prime\prime}=
-2\mathcal{F}^{(2)\prime}\cdot\mathcal{H}^\prime\cdot\mathcal{H}^{-1}+\sigma\mathcal{F}^{(1)}\,,\quad
\ldots\,,
\label{eq:born1}
\end{equation}
with boundary conditions $\lim_{x\to\infty}\mathcal{F}^{(n)}(x)=0$. In turn this Born expansion 
yields the $n^{\rm th}$ order contribution to the
phase shift $\delta_\ell^{(n)}(k)$ by collecting terms of order $n$ in 
Eq.~(\ref{eq:phase}). Using Eq.~(\ref{eq:vpe1}), each order can be associated with an energy 
computed from scattering data. Hence we have two expansions with respect to the same quantity.
They are equal order by order and we can write
\begin{align}
\Delta E=\frac{1}{2}\sum_{j}^{\rm b.s.}\omega_j
+\int_0^\infty \frac{dk}{2\pi} \sum_\ell
\sqrt{k^2+m^2}\,\left[\frac{d\delta_\ell(k)}{dk}\right]_N
+\sum_{n=1}^N E_{\rm FD}^{(n)}+E_{\rm CT}\,,
\label{eq:almostmaster}
\end{align}
where the subscript indicates that the first $N$ terms of the expansions are subtracted:
$\left[\delta_\ell(k)\right]_N=\delta_\ell(k)-\sum_{n=1}^N\delta_\ell^{(n)}(k)$. Integrating 
by parts and using Levinson's theorem \cite{Barton}
yields the VPE in terms of binding energies as
\begin{align}
\Delta E=\frac{1}{2}\sum_{j}^{\rm b.s.}\left(\omega_j-m\right)
-\int_0^\infty \frac{dk}{2\pi}\sum_\ell
\frac{k}{\sqrt{k^2+m^2}}\,\left[\delta_\ell(k)\right]_N
+\sum_{n=1}^N \left(E_{\rm FD}^{(n)}+E_{\rm CT}\right)\,.
\label{eq:almostmaster1}
\end{align}
The power of this expression is that it is a combination of finite terms. The momentum 
integral has become finite because sufficiently many $\delta_\ell^{(n)}(k)$ are subtracted from the 
exact phase shift. The Feynman diagram and counterterm energies are computed with a common
regularization scheme such that the regulator disappears. Eventually we will use dimensional
regularization with on-shell renormalization conditions; {\it i.e.} poles and residues of 
propagators do not have quantum corrections. These conditions are augmented by the ``no-tadpole''
condition that quantum corrections to the vacuum expectation values of the fields vanish.

We stress that for the formalism to be valid, the convergence of the expansion in 
Eq.~(\ref{eq:fdseries}) is not necessary because we only consider a finite number of terms.
It is also important to stress that, though the Born series features prominently in the 
approach, the phase shifts are not obtained by any kind of approximation.

We have obtained Eq.~(\ref{eq:almostmaster1}) by relating the change in the density of states
to the phase shift. The quantum field theory derivation that integrates the vacuum expectation 
value of the energy density can be found in Ref.~\cite{Graham:2002xq}.

\subsection{Imaginary momenta}
\label{ssec:IM}
It will be helpful in our calculations to make use of the analytic
properties of scattering data. Note that
$\mathcal{F}_{\ell,k}(x)\cdot\mathcal{H}_{\ell,k}(x)$ is the Jost solution to the scattering problem,
while $F_\ell(k)=\lim_{x\to0}{\rm det}\mathcal{F}_{\ell,k}(x)$ is the Jost function, which is 
analytic for ${\sf Im}(k)\ge0$. For the definition in terms of outgoing waves, the elements of 
$\mathcal{H}_{\ell,k}(x)$ are Hankel functions with asymptotic behavior proportional 
to ${\rm e}^{\imu kx}$. It is therefore clear that $\mathcal{F}_{-k}(x)=\mathcal{F}^\ast_k(x)$ 
for real~$k$. We then observe that the phase shift
\begin{equation}
\delta_\ell(k)=\frac{\imu}{2}\left(\ln F_\ell(k) -\ln F_\ell(-k)\right)
\label{eq:phase1}
\end{equation}
is an odd function of $k$. We thus write
\begin{equation}
\int_0^\infty \frac{dk}{2\pi}\sum_\ell
\sqrt{k^2+m^2}\,\left[\frac{d\delta_\ell(k)}{dk}\right]_N
=\frac{\imu}{2}\int_{-\infty}^\infty \frac{dk}{2\pi}
\sum_\ell \sqrt{k^2+m^2}\,
\left[\frac{d \ln F_\ell(k)}{dk}\right]_N\,.
\label{eq:acont1}
\end{equation}
To evaluate the right-hand-side by analytic continuation and contour integration, we recall that
\begin{itemize}
\item[$\bullet$]
with the subtraction of the Born terms, there is no contribution from the 
semicircles at $|k|\to\infty$;
\item[$\bullet$]
the Jost function has simple zeros at the bound state wave numbers while the Born 
terms have no zeros (otherwise bound states would be perturbative effects):
$$
\left[\frac{d \ln F_\ell(k)}{dk}\right]_N=\frac{1}{k-\imu\kappa_j}+\ldots
\qquad {\rm for}\qquad k\approx \imu\kappa_j\,;
$$
\item[$\bullet$]
the square root induces a branch cut along the imaginary axis $k=\imu t$:
$$
\sqrt{(\imu t+\epsilon)^2+m^2}-\sqrt{(\imu t-\epsilon)^2+m^2}=2\imu\sqrt{t^2-m^2}
$$
for $t\ge m$ and $\epsilon\to0^+$.
\end{itemize}
Collecting pieces we have
\begin{align}
&\int_0^\infty \frac{dk}{2\pi}\sum_\ell
\sqrt{k^2+m^2}\,\left[\frac{d\delta_\ell(k)}{dk}\right]_N\cr
&\hspace{2cm}
=-\frac{1}{2}\sum_j^{\rm b.s.}\sqrt{m^2-\kappa_j^2}
-\int_m^\infty \frac{dt}{2\pi}\sum_\ell
\sqrt{t^2-m^2}\left[\frac{d\ln F_\ell(it)}{dt}\right]_N\,,
\label{eq:acont2}
\end{align}
where the sum on the right-hand-side cancels the explicit bound state
contribution in Eq.~(\ref{eq:almostmaster})
\cite{Bordag:1996fv}. Again integrating by parts yields for the VPE
\begin{equation}
\Delta E=\int_m^\infty \frac{dt}{2\pi}\sum_\ell
\frac{t}{\sqrt{t^2-m^2}}\left[\nu_\ell(t)\right]_N
+\sum_{n=1}^N E_{\rm FD}^{(n)}+E_{\rm CT}\,,
\label{eq:master}
\end{equation}
where $\nu_\ell(t)=\ln F_\ell(\imu t)=\lim_{x\to0}\ln{\rm det}\mathcal{F}_{\ell,\imu t}(x)$.
We obtain $\mathcal{F}_{\ell,\imu t}(x)$ and its Born expansion by integrating Eqs.~(\ref{eq:SE2})
and~(\ref{eq:born1})  with $\mathcal{H}_{\ell,k}(x)\,\to\,\mathcal{H}_{\imu t}(x)$. It is 
worth noting that ${\rm det}\mathcal{F}_{\ell,\imu t}(x)$ is a real quantity.

In certain cases the limit $x\to0$ needs specific treatment. In two space dimensions, the 
regular solution approaches a constant while the irregular one diverges logarithmically 
in the zero angular momentum channel. Numerically these solutions
 are cumbersome to disentangle and it
is not sufficient simply to read off $\mathcal{F}_{\ell,\imu t}(x)$ at some very small $x$; rather a 
sophisticated extrapolation is required \cite{Graham:2011fw}. Furthermore singular background 
potentials may require additional subtractions because the Born expansion reflects 
that singularity, for example in the case of vortices in scalar electrodynamics \cite{Graham:2019fzo}.
Also, fermion masses explicitly enter the spinor wave-functions and the effective masses at
spatial infinity and $x\to0$ may differ. This effect must be incorporated when extracting 
$\nu(t)$ from $\mathcal{F}_{\ell,\imu t}(x)$ as well \cite{Graham:2011fw}. We will return
to these issues in Secs.~\ref{sec:cstring} and~\ref{sec:vortices}.

\subsection{Interface configurations}
\label{ss:interface}
A problem that often arises in contexts such the Casimir
force \cite{Graham:2002xq} is that of computing the VPE originating from (idealized) boundary 
conditions of infinitely large plates. In such a problem, the plate acts like a domain wall, 
with translational invariance along a plane, while there is a localized structure in the
orthogonal dimension. Similarly, strings stretch along an axis with translational invariance
while having a soliton-like configuration in the plane orthogonal to that axis. The scattering 
problem in the translationally invariant subspace is trivial, which introduces subtleties 
in the computation of the VPE and in particular its regularization and renormalization.

We consider a general scenario with a
localized (classical) configuration contained in the $n$-dimensional
subspace $\mathbb{R}^n$.
It is embedded in a higher-dimensional space such that there is translational invariance in an 
$m$-dimensional subspace $\mathbb{R}^m$. The combined fluctuation
wavefunction is then
\begin{equation}
\varphi(\Vek{x},\Vek{y})= {\rm e}^{\imu\vek{p}\cdot\vek{y}}\,
\tilde{\varphi}(\Vek{x})\,,
\label{eq:interfwfct}
\end{equation}
where $\Vek{p}\in\mathbb{R}^m$ is the transverse momentum conjugate to the extra
dimensions $\Vek{y}$. The reduced wave function $\widetilde{\varphi}(\Vek{x})$ is
subject to a wave-equation in $n$ spatial dimensions with a background 
potential $\sigma(x)$ similar to Eq.~(\ref{eq:SE1}). Then the
dispersion relations for the scattering and bound states of the $n+m$
dimensional problem are ($p=|\Vek{p}|$ and $k=|\Vek{k}|$)
$$
\omega(k,p) = \sqrt{m^2+k^2+p^2}
\qquad {\rm and} \qquad
\omega_{j}(p)=\sqrt{m^2+p^2-\kappa^2_{j}}\,,
$$
respectively. Here $k$ and $\kappa_j$ are the (imaginary) momenta conjugate to 
$\Vek{x}\in\mathbb{R}^n$. We get the corresponding densities of states
by simply multiplying the free density of states of the $m$
dimensional transverse space
\begin{equation}
\rho_\ell(\Vek{p},k)=\frac{1}{\pi}\frac{d\delta_\ell(k)}{dk}\frac{V_m}{(2\pi)^m} \qquad {\rm and} \qquad
\rho_{j}(\Vek{p})=\frac{V_m}{(2\pi)^m}\,.
\end{equation}
This in turn yields the VPE per unit transverse volume
\begin{align}
\mathsf{E}_{m,n} = \frac{\Delta E}{V_m} &=
\int \frac{d^m p}{(2\pi)^m}
\biggl[\int_0^\infty \frac{dk}{2\pi}  \sum_\ell
\left(\omega(k,p)-\mu(p)\right) \frac{d\delta_\ell(k)}{dk} \nonumber \\[2mm]
&\hspace{1cm}\qquad \quad
+\frac{1}{2}\sum_j \left(\omega_{j}(p)-\mu(p)\right)\biggr] 
+ \frac{E_{\rm CT}}{V_m}\,,
\label{eq:regularized}
\end{align}
where the mass coefficient $\mu(p)=\sqrt{m^2+p^2}$ must be taken to depend on the 
transverse momentum. In this form the subtlety mentioned above emerges immediately: 
The $p$ integral is divergent but the phase shift does not involve $p$ and thus there 
is no Born subtraction that can render this integral finite. To see this more clearly, 
we integrate over $p$ in dimensional regularization:
\begin{align}
\mathsf{E}_{m,n}& = -\frac{\Gamma(-\frac{1+m}{2})}{2(4\pi)^{\frac{m+1}{2}}}
\left[\int_0^\infty \frac{dk}{\pi} \sum_\ell
(\omega^{m+1}(k,0) - m^{m+1})\,
\frac{d}{dk}\left[\delta_\ell(k) \right]_N  + \right.
\nonumber \\[2mm]
& \hspace{2.5cm}
+\sum_j (\omega_{j}(0)^{m+1}-m^{m+1}) \biggr] + 
\frac{E^{(N)}_{\rm FD}}{V_m}+\frac{E_{\rm CT}}{V_m}\,,
\label{eq:inter0}
\end{align}
where we have subtracted $N$ Born terms from the phase shift and added them back as the 
Feynman diagrams $E^{(N)}_{\rm FD}$. Regardless of the number of subtractions, the 
coefficient produces a divergence for any odd integer $m$. Let us specifically consider
the case $m=1$, which applies to string configurations. Then the
expression in square brackets simplifies to
$$
\int_0^\infty \frac{dk}{\pi} k^2 \frac{d}{dk}\left[\delta_\ell(k) \right]_N-\sum_j \kappa_j^2
=\int_0^\infty \frac{dk}{\pi}\,
\omega^2(k)\frac{d}{dk}\left[\delta_\ell(k) \right]_N
+\sum_j\omega_j^2 \,,
$$
where the energies refer to those in the $n$-dimensional subspace. The equality above
is a consequence of Levinson's theorem. More importantly, this combination vanishes
by one of the sum rules that generalize that theorem \cite{Graham:2001iv}. Hence the 
residue of the pole from $\Gamma(-\frac{1+m}{2})$ is zero and we can analytically continue 
to $m=1$,
\begin{align}
\mathsf{E}_{1,n}& =
\frac{-1}{8\pi} \left[\int_0^\infty \frac{dk}{\pi}\sum_\ell \omega^2(k)
\ln\frac{\omega^2(k)}{\overline{\mu}^2}\,
\frac{d}{dk}\left[\delta_\ell(k) \right]_N
+ \sum_j \omega_{j}^2\ln\frac{\omega_{j}^2}{\overline{\mu}^2}\right]
+\frac{E^{(N)}_{\rm FD}}{V_m}+\frac{E_{\rm CT}}{V_m}\,.
\label{eq:inter1}
\end{align}
The arbitrary energy scale $\overline{\mu}$ has been introduced for dimensional reasons. 
It has no effect by the sum rule mentioned above.
Making use of the relation between the phase shift and the Jost function and its analytic properties 
allows us to write the integral for imaginary momenta as in Sec.~\ref{ssec:IM}. While the 
bound states energies will no longer appear explicitly, we pick up a contribution from the 
discontinuity of the logarithm and integrate by parts:
\begin{align}
\mathsf{E}_{1,n}=\int_m^\infty\frac{dt}{4 \pi}\, t\, \left[ \nu(t) \right]_N
+\frac{E^{(N)}_{\rm FD}}{V_m}+\frac{E_{\rm CT}}{V_m}\,,
\label{eq:inter2}
\end{align}
where $\nu(t)=\sum_\ell\nu_\ell(t)$ is the channel sum of the logarithms of the partial wave Jost functions 
in the $n$ dimensions that contain the soliton.

\subsection{Fake boson subtraction}
\label{ssec:fake}

In many cases the logarithmically divergent Feynman diagrams are cumbersome to compute.
An example is the fermion loop in $D=3+1$ dimensions, which requires considering
diagrams with up to four insertions of the background potential. It is thus desirable to 
have available a simpler treatment of the logarithmic divergences. The key observation is 
that the sole purpose of introducing the Born terms and the equivalent Feynman diagrams is 
to move the ultraviolet divergences out of the momentum integral and combine them with the 
counterterms, $E_{\rm CT}$.  Hence we can use any other subtraction under the momentum integral
that can be combined with the same counterterms, $E_{\rm CT}$. 
In dimensional regularization, $D=3+1\,\to\,4-2\epsilon$,
the logarithmic divergence emerges as $\frac{C_L}{32\pi^2\epsilon}$ where $C_L$ is a local
integral over (powers of) the potential in the wave equation. We also note that the Feynman diagram
with a scalar loop and two insertions of a potential $V_{\rm B}$ leads to a logarithmic divergence in 
$E_{\rm FD}^{(2)}$. This divergence is proportional to the spatial integral of $V^2_{\rm B}$. In this 
boson theory we can compute $\nu_{\rm B}^{(2)}(t)$, the channel sum of the second order (in $V_B$) 
contribution to the (summed) logarithm of the Jost function. Next we take $\overline{\nu(t)}$
to be the logarithm of the Jost function in the original theory with all Born terms removed that lead 
to divergences higher than logarithmic. This also subtracts the underlying logarithmic divergences in
the corresponding Feynman diagrams, so we only need to consider Feynman diagrams that have superficial
logarithmic divergences. We add back these subtractions as Feynman diagrams and combine them 
with the counterterms to $E_{\rm CT}$. We then consider
\begin{equation}
\int_m^\infty \frac{dt}{2\pi}\frac{t}{\sqrt{t^2-m^2}}
\left[\overline{\nu(t)}-C_{\rm B}\nu_{\rm B}^{(2)}(t)\right]
+\overline{E_{\rm FD}}+C_{\rm B}E_{\rm FD}^{(2)}+E_{\rm CT}\,,
\label{eq:fb1}
\end{equation}
where $\overline{E_{\rm FD}}$ are the Feynman diagrams compensating the subtractions in $\overline{\nu(t)}$.
Setting $C_{\rm B}=\frac{C_L}{\int d^3x\,V^2_{\rm B}}$ moves the logarithmic divergence from the 
momentum integral to the Feynman diagram such that 
$\overline{E_{\rm FD}}+C_{\rm B}E_{\rm FD}^{(2)}+E_{\rm CT}$ is finite.

As a corollary to this prescription we can consider the finite differences between second-order terms in 
boson theories with different potentials $V_1$ and $V_2$ that are related by the integrals
$\int d^3x\, V_1^2=\int d^3x\, V_2^2$. They have second-order phase shifts $\delta_{\ell,1}^{(2)}(k)$ and 
$\delta_{\ell,2}^{(2)}(k)$, respectively.
The procedure in Eq.~(\ref{eq:fb1}) implies the relation
(with no discrete contributions since Born terms do not account for
the non-perturbative bound states),
$$
-\int_0^\infty \frac{dk}{2\pi} \sum_\ell
\frac{k}{\sqrt{k^2+m^2}}\left[\delta_{\ell,1}^{(2)}(k) -
\delta_{\ell,2}^{(2)}(k)\right]=E_{{\rm FD},1}-E_{{\rm FD},2}\,,
$$
which has been numerically verified for particular examples
\cite{Farhi:2001kh}.

Finally we note that this simple {\it fake boson} subtraction does not
work for quadratic divergences, because then the underlying logarithmic
divergences must also be accounted for.

\section{Applications in one space dimension}
\label{sec:Dim1}

In this Section we demonstrate the effectiveness of the imaginary
momentum formulation by computing the VPE of sine-Gordon and kink
solitons. We also summarize some recent computations of
the VPEs for other solitons in one dimension.

\subsection{Exactly solvable examples}

In Sec.~\ref{ssec:IM} we have emphasized the convenience of the imaginary momentum
formalism. Let us briefly demonstrate this in the case of the
sine-Gordon and $\phi^4$ kink solitons,
for which the typical textbook calculations of the VPE are carried out 
using real momenta \cite{Ra82,Graham:2009zz}. First we note that for boson theories there 
is only one divergent Feynman diagram in $D=1+1$, which is the tadpole with a single insertion of
the background potential (first diagram in Eq.~(\ref{eq:fdseries})). This diagram can 
be fully canceled by a counterterm such that $E^{(1)}_{\rm FD}+E_{\rm CT}=0$. Accordingly we
only have a single Born subtraction under the momentum integral. The background potentials 
are of the P\"oschl-Teller form \cite{Poschl:1933zz}
\begin{equation}
\sigma_n(x)=-\frac{n+1}{n}\,\frac{M^2}{\cosh^2\left(\frac{Mx}{n}\right)}
\label{eq:PT1}
\end{equation}
with $n=1$ and $n=2$ for the sine-Gordon and $\phi^4$ kink solitons,
respectively.
In both cases $M$ is the mass parameter for the quantum fluctuations,
and the scattering solutions are known:
\begin{align}
\psi_k^{(1)}(x)&\propto {\rm e}^{\imu kx}\left[k+\imu M\tanh(Mx)\right]\\
\psi_k^{(2)}(x)&\propto {\rm e}^{\imu kx}\left[3\tanh^2\left(\frac{Mx}{2}\right)-1
-\frac{4k^2}{M^2}-\frac{6\imu k}{M}\tanh\left(\frac{Mx}{2}\right)\right]\,,
\label{eq:PT2}
\end{align}
where the superscript is the P\"oschl-Teller index $n$. The exponential
factor indicates that we have only
a right-moving plane wave. Hence the potentials are reflectionless and the phase shifts in the 
symmetric and anti-symmetric channels are equal. The Jost solution ($f=\mathcal{F}\mathcal{H}$) is 
constructed from the above wave-functions by introducing constant factors such that
$$
\lim_{x\to\infty}f_k^{(n)}(x)\,{\rm e}^{-\imu kx}=1\,.
$$
Since the wave-equation is real, both $f_k(x)$ and $f_k^{\ast}(x)$ are solutions and
the Jost functions are the expansion coefficients of $f$ and $f^\ast$
for the physical scattering solution 
$\psi_{\rm sc}$. In the antisymmetric channel the boundary conditions read
$$
\lim_{x\to0}\psi_{\rm sc}(x)=0 \qquad {\rm and}\qquad
\lim_{x\to0}\frac{d\psi_{\rm sc}(x)}{dx}=1\,.
$$
Equating the Wronskian $W\left[\psi_{\rm sc}(x),f_k^{(n)}(x)\right]$ at spatial infinity
and at $x\to0$ immediately yields the Jost function $F_{-}(k)=\lim_{x\to0}f_k(x)$ 
as denoted at the end of Sec.~\ref{ssec:SD}. For the above potentials this gives
\begin{equation}
F_{-}^{(1)}(k)=\frac{k}{k+\imu M} \qquad {\rm and}\qquad
F_{-}^{(2)}(k)=\frac{k-\imu M/2}{k+\imu M}\,.
\label{eq:PT3}
\end{equation}
As mentioned in footnote \ref{ft:sym}, the boundary condition in the symmetric channel requires
the derivative of the Jost solution such that $F_{+}(k)=\frac{1}{\imu k}\lim_{x\to0}\frac{df_k(x)}{dx}$,
yielding
\begin{equation}
F_{+}^{(1)}(k)=\frac{k-\imu M}{k} \qquad {\rm and}\qquad
F_{+}^{(2)}(k)=\frac{k-\imu M}{k+\imu M/2}\,.
\label{eq:PT4}
\end{equation}
We note that $F_{+}^{(1,2)}(\imu M)=0$ reflects the existence of zero modes. The analytic 
continuation for $F^{(1,2)}_{+}F^{(1,2)}_{-}$ is straightforward, leading to the VPE
\begin{align}
\Delta E^{(1)}&=\int_M^\infty \frac{dt}{2\pi} \frac{t}{\sqrt{t^2-M^2}}
\left[\ln\left(\frac{t-M}{t+M}\right)+2\frac{M}{t}\right]=-\frac{M}{\pi}
\label{eq:PT5} \\
\Delta E^{(2)}&=\int_M^\infty \frac{dt}{2\pi} \frac{t}{\sqrt{t^2-M^2}}
\left[\ln\left(\frac{t-M}{t+M}\frac{2t-M}{2t+M}\right)+3\frac{M}{t}\right]
=\frac{M}{12}\left(\sqrt{3}-\frac{18}{\pi}\right)\,.
\nonumber \end{align}
In the above we have identified the Born subtraction as the leading term
of the logarithm for large $t$. From the wave-equation we can show that it is 
indeed a single inverse power in $t$. We factorize
$f_k(x)=\mathcal{F}_{\ell,k}(x){\rm e}^{-\imu kx}$ with the imaginary
momentum wave-equation (primes are derivatives with respect to the
spatial coordinate $x$)
$$
\mathcal{F}^{\prime\prime}_{\imu t}(x)=2t\mathcal{F}^\prime_{\imu t}(x)
+\sigma(x)\mathcal{F}_{\ell,\imu t}(x)
$$
and obtain the full Jost function (argument of the logarithms in Eqs.~(\ref{eq:PT5})) as 
the limit
\begin{equation}
F_{+}(t)F_{-}(t)=\lim_{x\to0}\left[\mathcal{F}_{\ell,\imu t}(x)\left(
\mathcal{F}_{\ell,\imu t}(x)-\frac{1}{t}\mathcal{F}^\prime_{\imu t}(x)\right)\right]\,.
\label{eq:full1D}
\end{equation}
To identify the Born approximation we
expand $\mathcal{F}_{\ell,\imu t}(x)=1+\mathcal{F}_1(x,t)+\ldots$ in powers of the 
potential $\sigma$. Integrating the differential equation for $\mathcal{F}_1(x,t)$
from the center to infinity shows that
\begin{equation}
\ln\left[F_{+}(t)F_{-}(t)\right]
=\frac{1}{t}\int_0^\infty dx\, \sigma(x)+\mathcal{O}\left(\sigma^2\right)\,.
\label{eq:born1D} \end{equation}

\subsection{Mass gap and thresholds}
\label{subs:gap}

The imaginary momentum formulation is even more advantageous when there are
multiple quantum fields with different masses. In the real momentum formulation,
three different energy regimes must be considered: 1) bound 
states with energies below the smallest mass; 2) intermediate energy regime(s) between
the lowest and largest masses where some modes are bound while others scatter;
3) energies above the largest mass, where all modes scatter and extend
to spatial infinity.

The continuation to imaginary momenta, however, is not without obstacles.
Let us explore the case with two masses $m_1<m_2$ that have momenta $k_1=k$ and
$k_2$ respectively. In the binding regime(s) they may be imaginary. For static
potentials, energy is conserved and the (relativistic) dispersion relation
yields $k_2^2=k^2+m_1^2-m_2^2$. This may induce further branch cuts in the 
complex momentum plane. Within the mass gap, $0<k^2<m_2^2-m_1^2$, either sign of $k$ 
should produce an exponentially decaying wave-function parameterized by the 
imaginary part of $k_2$. This suggests writing $k_2=\sqrt{k^2+m_1^2-m_2^2}$. 
In the scattering regime, $k^2>m_2^2-m_1^2$, the continuation uses the fact that 
the imaginary part of the Jost function is odd for real momenta, as
shown in Eq.~(\ref{eq:SE1}). In particular, $k_2$ should change sign when $k$ does. So 
we would write $k_2=k\sqrt{1+\frac{m_1^2-m_2^2}{k^2}}$. In Ref.\cite{Weigel:2017kgy}
it has been shown that these two seemingly contradictory relations can be
consistently combined as
\begin{equation}
k_2=k_2(k)\equiv k\sqrt{1+\frac{m_1^2-m_2^2}{\left[k+\imu\epsilon\right]^2}}
\qquad {\rm with}\qquad \epsilon\to0^+\,.
\label{eq:k2}\end{equation}
This prescription leads to additional singularities, but they occur
only for momenta with 
negative imaginary parts, and it passes numerous consistency checks \cite{Weigel:2017kgy}. 
Hence it is straightforwardly permissible to analytically continue within the 
upper half plane with $k=\imu t$ and $k_2=\imu t_2(t)=\imu \sqrt{t^2+m_2^2-m_1^2}$. 
We then get the matrix differential equation
\begin{equation}
\mathcal{F}_{\ell,\imu t}^{\prime\prime}(x)=
2\mathcal{F}_{\ell,\imu t}^\prime(x)D(t)+\left[M_2^2,\mathcal{F}_{\ell,\imu t}(x)\right]+
V(x)\mathcal{F}_{\ell,\imu t}(x)\,,
\label{eq:masterD1} \end{equation}
with $M_2=\begin{pmatrix} m_1 & 0 \cr 0 & m_2\end{pmatrix}$
and $D(t)=\begin{pmatrix} t & 0 \cr 0 & \,t_2\end{pmatrix}$.

If the potential matrix is symmetric, $V(-x)=V(x)$, the relevant logarithm 
of the Jost functions is
\begin{equation}
\nu(t)={\rm ln} \, {\rm det}\left[F_{S}(t)F_{A}(t)\right]\,,
\label{eq:nuSD1} \end{equation}
where $F_{S}(t)=\lim_{x\to0}\left[\mathcal{F}_{\ell,\imu t}(x)-\mathcal{F}_{\ell,\imu t}^\prime(x)D^{-1}(t)\right]$
and $F_{A}(t)=\lim_{x\to0}\left[\mathcal{F}_{\ell,\imu t}(x)\right]$. In many applications
the potential matrix has a skewed symmetry
$V(-x)=\begin{pmatrix}1 & 0 \cr 0 & -1\end{pmatrix}\, V(x)\, 
\begin{pmatrix}1 & 0 \cr 0 & -1\end{pmatrix}$,
in which case \cite{Weigel:2017kgy,Weigel:2018jgq}
\begin{equation}
\nu(t) = {\rm ln}\,{\rm det}\left[F_{+}(t) \,F_{-}(t)\right]\,,
\label{eq:nuSKD1} \end{equation} 
where the combinations
$F_{\pm}(t)=\left[P_{\pm}F_S(t)D_{\mp}(t)+P_{\mp}F_A(t)D_{\pm}^{-1}(t)\right]$
are found using the projectors
$P_{+}=\begin{pmatrix}1 & 0 \cr 0 & 0\end{pmatrix}$
and
$P_{-}=\begin{pmatrix}0 & 0 \cr 0 & 1\end{pmatrix}$
as well as the factor matrices
$D_{+}(t)=\begin{pmatrix}-t & 0 \cr 0 & 1\end{pmatrix}$
and
$D_{-}(t)=\begin{pmatrix}1 & 0 \cr 0 & -t_2\end{pmatrix}$. The Born approximation,
\begin{equation}
\nu^{(1)}(t)=\int_{0}^\infty dx\, \left[\frac{V_{11}(x)}{t}
+\frac{V_{22}(x)}{t_2}\right]\,,
\label{eq:Born1}\end{equation}
is then subtracted from Eq.~(\ref{eq:nuSKD1}) to implement the
no-tadpole scheme.

\subsection{Instability of Shifman-Voloshin soliton}

The Shifman-Voloshin soliton model extends the $\phi^4$ model 
by adding a second scalar field $\chi$. Its Lagrangian reads
\begin{equation}
\mathcal{L}=\frac{1}{2}\left[\partial_\nu \phi\partial^\nu \phi
+\partial_\nu \chi\partial^\nu \chi\right]
-\frac{\lambda}{4}\left[\phi^2-\frac{M^2}{2\lambda}
+\frac{\mu}{2}\chi^2\right]^2-\frac{\lambda}{4}\mu^2\chi^2\phi^2\,.
\label{eq:Bazeia}\end{equation}
The Lagrangian contains a coupling constant $\lambda$ and mass scale $M$ similar to the 
conventional $\phi^4$ model. We will discuss the meaning of the dimensionless coupling
constant $\mu>0$ shortly.

After appropriate redefinition of the fields, $(\phi,\chi)\to(M/\sqrt{2\lambda})(\phi,\chi)$ 
and the coordinates, $x_\nu\to2x_\nu/M$, the rescaled Lagrangian,
$\mathcal{L}\to(M^4/8\lambda)\mathcal{L}$ has a vacuum configuration at $\phi=\pm1$ and $\chi=0$. 
We call these two possibilities the primary vacua. There are also secondary vacua at 
$\phi=0$ and $\chi=\pm\sqrt{\frac{2}{\mu}}$. 

For static fields the model allows a BPS construction for the classical energy of static fields
\begin{align}
E_{\rm cl}&=\frac{1}{2}\int_{-\infty}^\infty dx\,
\left[\phi^{\prime2}+\chi^{\prime2} +\left(\phi^2-1+\frac{\mu}{2}\chi^2\right)^2
+\mu^2\phi^2\chi^2\right]
\label{eq:bps}\\
&=\frac{1}{2}\int_{-\infty}^\infty dx\,
\left[\left(\phi^2-1+\frac{\mu}{2}\chi^2\pm\phi^\prime\right)^2
+\left(\mu\phi\chi\pm\chi^\prime\right)^2\right]
\pm\left[\phi-\frac{1}{3}\phi^3-\mu\phi\chi^2\right]_{-\infty}^\infty\,,
\nonumber \end{align}
where prime denotes a derivative with respect to the (dimensionless) space
coordinate $x$. The extremal points in field space are determined from the
first-order differential equations
\begin{equation}
\frac{d\chi(x)}{dx}=-\mu\phi(x)\chi(x)
\qquad{\rm and}\qquad
\frac{d\phi(x)}{dx}=1-\phi^2(x)-\frac{\mu}{2}\chi^2(x)\,.
\label{eq:lindeq}\end{equation}
These coupled differential equations have been studied in detail in 
Refs.~\cite{Shifman:1997wg,Bazeia:1995en}. Field configurations that approach the 
secondary vacuum at spatial infinity have $E_{\rm cl}=0$, and we thus cannot have 
a soliton with this asymptotic behavior. Adopting the convention that 
$\lim_{x\to\pm\infty}\phi(x)=\pm1$, we see that $\phi(x)$ is an increasing
(presumably monotonically) odd function of the coordinate and $\chi$ is
even, where we take the soliton center at $x=0$. We are free to choose $\chi(0)\ge0$. 
If $\chi(0)>\sqrt{2/\mu}$, $\phi^\prime(0)<0$ and $\chi^{\prime\prime}(0)>0$, implying 
that $\chi(0)$ would be a minimum. Furthermore $\phi^\prime$ would turn
even more negative and not approach $+1$ at spatial infinity. By contradiction
we thus conclude that $\sqrt{2/\mu}$ is an upper bound for $\chi(0)$
and we parameterize $\chi(0)=a\sqrt{2/\mu}$ with $0\le a<1$, for which solitons
have been constructed numerically \cite{Weigel:2018jgq}. Observe that $\phi$ is 
in its secondary vacuum at $x=0$. The closer $a$ is to unity, the larger the region in 
which both fields approximately equal their corresponding expectation values from 
the secondary vacuum. Restoring units we see from Eq.~(\ref{eq:bps}) that 
$E_{\rm cl}=\frac{4}{3}\frac{M^3}{8\lambda}=\frac{M^3}{6\lambda}$ for all solitons, 
independent of $a$.

\begin{figure}[ht]
  \centerline{\includegraphics[width=6.2cm,height=4cm]{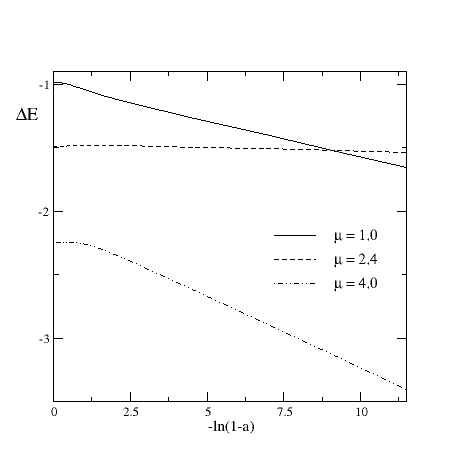}~~~
              \includegraphics[width=6.2cm,height=4cm]{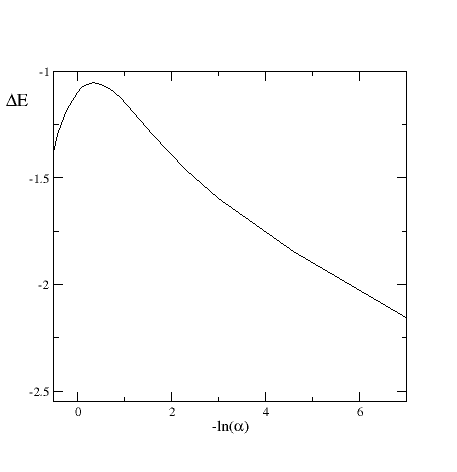}~~}
  \caption{VPE of the Shifman-Voloshin soliton for various model
parameters $\mu$ as functions of the variational parameter $a$ (left panel)
and the $\phi^6$ model soliton as a function of the
model parameter $\alpha$ (right panel).}
\label{fig:vpe}
\end{figure}

Linearizing the time-dependent wave-equations around this soliton defines the 
potential and mass matrices
\begin{equation}
V(x) =\begin{pmatrix}
\mu(1+\mu)\left(\phi^2-1\right)+\frac{3}{2}\mu^2\chi^2 &
2\mu(1+\mu)\chi\phi \\[2mm] 2\mu(1+\mu)\chi\phi &
6\phi^2-6+\mu(\mu+1)\chi^2
\end{pmatrix}
\quad {\rm and}\quad
M_2=\begin{pmatrix}\mu & 0 \cr 0 & 2\end{pmatrix}\,,
\label{eq:vpot}
\end{equation}
respectively. Obviously $V(x)$ is skew-symmetric and for $\mu\le2$ we can directly 
apply the formalism of Sec.~\ref{subs:gap}, while for $\mu>2$ we first need to 
swap the diagonal elements of both $V(x)$ and $M_2$. Selected results for the 
VPE from Ref. \cite{Weigel:2018jgq} are listed in 
Tab.~\ref{tab:vpeSV}.
\begin{table}[htbp]
\tbl{Numerical results for the vacuum polarization energy
of the Shifman-Voloshin soliton, measured in units of $M$.}
{\begin{tabular}{l|ccccccc}
%\hline
\diagbox[height=0.8cm]{~~$a$}{$\mu$} & 0.5  & 1.6
& 2.0 & 2.8 & 3.6 & 4.0 & 4.4 \cr
\hline
0.0     & -0.830 & -1.186 & -1.333 & -1.661 & -2.039 & -2.246 & -2.467\cr
0.1     & -0.833 & -1.186 & -1.333 & -1.661 & -2.038 & -2.246 & -2.467\cr
0.5     & -0.906 & -1.195 & -1.333 & -1.654 & -2.036 & -2.249 & -2.477\cr
0.9     & -1.229 & -1.217 & -1.333 & -1.666 & -2.112 & -2.372 & -2.656\cr
0.99    & -1.661 & -1.235 & -1.333 & -1.714 & -2.284 & -2.628 & -3.008\cr
0.999   & -2.076 & -1.251 & -1.333 & -1.764 & -2.459 & -2.888 & -3.364\cr
0.9999  & -2.488 & -1.268 & -1.333 & -1.813 & -2.634 & -3.147 & -3.720\cr
0.99999 & -2.900 & -1.284 & -1.333 & -1.863 & -2.809 & -3.406 & -4.076
\end{tabular}\label{tab:vpeSV}}
\end{table}
Except for $\mu=2$, where the model is equivalent to two identical $\phi^4$ kinks,
the VPE is unbounded from below as $a\to1$. A numerical fit exhibits a logarithmic
divergence: $\Delta E\sim E_0+E_1{\rm ln}(1-a)$, with $E_{1,2}$ approximately 
independent of $a$. This behavior is shown in the left panel of Fig.~\ref{fig:vpe}.

Hence for any given $\lambda$, there will be an $a$ close to unity
such that the total energy is negative. By extending into the secondary vacuum 
the soliton is destabilized at the one-loop quantum level. For this to happen,
the curvatures of the field potential at the primary and the (degenerate) secondary 
vacua must be different; recall that for $\mu=2$ they are equal, and
then there is no instability.

\subsection{Further quantum instabilities: higher power field potentials}

\begin{figure}[t]
  \centerline{\includegraphics[width=6.2cm,height=4.5cm]{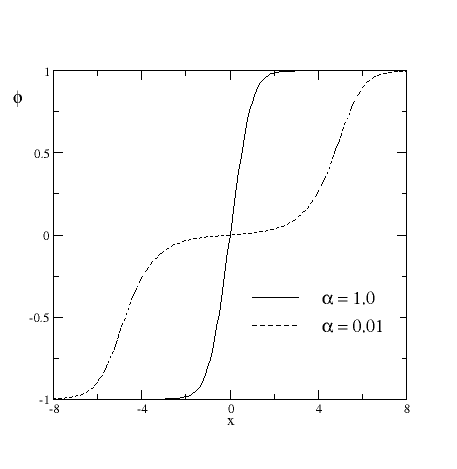}~~~
              \includegraphics[width=6.2cm,height=4.5cm]{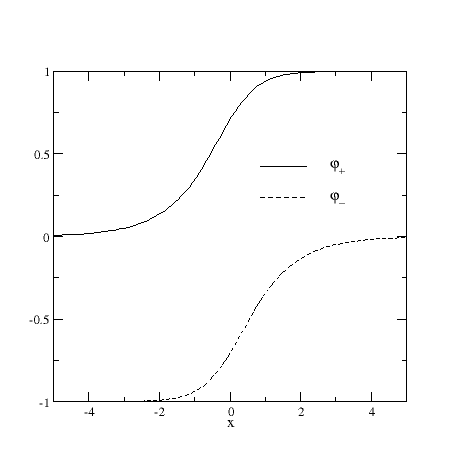}~}
  \caption{\label{fig:sol6}The soliton profiles of the $\varphi^6$ model. Note that
$\alpha=0$ in the right panel.}
\end{figure}

Models with higher than quartic powers in the field potential may also possess
primary and secondary vacua for certain parameters. An example is the Lagrangian
\begin{equation}
\mathcal{L}=\frac{1}{\lambda} \left[
\frac{1}{2}\partial_\nu\varphi\partial^\nu\varphi-U(\varphi)\right]
\qquad {\rm with} \qquad 
U(\varphi)=\frac{1}{2}\left(\varphi^2-1\right)^2\left(\varphi^2+\alpha^2\right)\,.
\label{eq:phi6}\end{equation}
For non-zero values of the model parameter $\alpha$, only primary vacua at $\varphi=\pm1$
exist. As shown in Fig.~\ref{fig:sol6}, solitons mediate between these two values as the 
spatial coordinate varies between negative and positive infinity \cite{Lohe:1979mh,Lohe:1980js}.
The potential for the harmonic fluctuations around this soliton is invariant under spatial 
reflection and the VPE can be straightforwardly computed using the methods described above, 
via Eq.~(\ref{eq:full1D}). 
The right panel in Fig.~\ref{fig:vpe} shows that for $\alpha\to0$, the VPE approaches negative 
infinity like ${\rm ln}\alpha$ \cite{Weigel:2016zbs,Alonso-Izquierdo:2011hmo}. On the other hand, 
the classical energy is always finite and positive. Classical and quantum energies contribute 
with different powers of the loop counting parameter $\lambda$, which
appears as an overall factor in $\mathcal{L}$. 
Hence there will always be a value of $\lambda$ such that the total energy is positive and the 
soliton is stable.
The situation changes drastically for $\alpha=0$. Then a secondary vacuum emerges at
$\varphi=0$ and two soliton solutions $\varphi_{\pm}$ exist that link $\varphi=0$ with 
$\varphi=1$ and $\varphi=-1$ with $\varphi=0$, respectively. Actually for a tiny but non-zero
$\alpha$, the soliton from above can be viewed as a combination of these two solutions whose
separation increases as $\alpha\to0$. This may also be inferred from the profile functions
shown in Fig.~\ref{fig:sol6}. 

Obviously the curvatures of the field potential at $\varphi=0$ and $\varphi=\pm1$ differ. 
Hence the masses of the quantum fluctuations at positive and negative infinity are also different. 
Even though an imaginary momentum formalism has not (yet) been developed for this scenario, 
the reflection and transmission coefficients as well as the bound state energies have been 
computed both analytically \cite{Lohe:1979mh} and numerically \cite{Weigel:2016zbs} 
so that the VPE can be obtained from Eq.~(\ref{eq:almostmaster1}). That simulation reveals a 
translational dependence. With $x_0$ being the center of the soliton, the VPE is observed to 
change by approximately $0.101$ per unit of $x_0$. The same variation is found for the VPE of 
a background potential built from a barrier of width $x_0$ and a height determined from the 
difference of the two curvatures. In contrast to $\alpha$, $x_0$ is a variational parameter 
that describes the shape of the soliton and for any given $\lambda$ we can choose $x_0$ such 
that the total energy is negative. Hence the $\alpha=0$ soliton is unstable at one-loop order. 
Of course, this picture is consistent with the $\alpha\to0$ divergence of the VPE discussed 
above, when the soliton configuration exhibits two well separated $\alpha\equiv0$ structures.
The same kind of instability has been observed in the $\varphi^8$ model \cite{Takyi:2020tvl}.

\section{Fermions and cosmic strings}
\label{sec:cstring}
In contrast to the Abelian vortices that we will discuss in Sec.~\ref{sec:vortices},
string configurations that are embedded in $SU(2)$ Higgs-gauge theory
are not stable by their topological 
structure, so it is of interest to explore whether quantum effects may lead to 
stabilization. For large $N$ (the number of internal fermion degrees
of freedom in the field theory) the fermion contribution will be dominant.

For the current investigation the fermion doublet will be assumed degenerate, so that 
the introduction of a matrix notation for the Higgs field is appropriate. In general the
isospin group $SU(2)$ is described by three Euler angles. One angle picks up the winding 
of the string in azimuthal direction. So we are left with two angles $\xi_1$ and $\xi_2$ 
that parameterize the isospin orientation\cite{Graham:2006qt}. They will later be treated as 
variational parameters. For notational simplicity, we introduce the abbreviations 
$s_i={\rm sin}(\xi_i)$ and $c_i={\rm cos}(\xi_i)$. Then the string configuration reads
\begin{equation}
\Phi=\begin{pmatrix}
\phi_0^* & \phi_+ \cr -\phi_+^* & \phi_0 \end{pmatrix}
=vf_H(\rho)
\begin{pmatrix}
s_1 s_2\, {\rm e}^{-in\varphi} & -ic_1-s_1c_2 \\[2mm]
-ic_1+s_1c_2 & s_1 s_2\, {\rm e}^{in\varphi}
\end{pmatrix}
\label{eq:Higgs}
\end{equation} 
for the Higgs field and 
\begin{equation}
\vek{W} = n\,s_1\,s_2\, \frac{f_G(\rho)}{g\rho}\,\hat{\Vek{\varphi}}\,
\begin{pmatrix}
s_1 s_2 & -\left(ic_1+s_1c_2\right){\rm e}^{in\varphi} \\[2mm]
\left(ic_1-s_1c_2\right){\rm e}^{-in\varphi} & -s_1s_2
\end{pmatrix}
\label{eq:gauge}
\end{equation}
for the gauge boson (in temporal gauge with $W_0 = 0$).
The variables $\rho$ and $\varphi$ are polar coordinates in the plane perpendicular 
to the string, while the gauge coupling constant $g$ and the Higgs vacuum expectation 
value~$v$ are model parameters. The profile functions $f_H$ and $f_G$ are subject
to the boundary conditions
\begin{equation}
f_G\,,\,f_H\, \to\, 0 
\quad {\rm for} \quad \rho\,\to\, 0
\quad {\rm and}\quad
f_G\,,\,f_H\, \to\, 1
\quad {\rm for} \quad \rho\,\to\, \infty\,.
\label{eq:bc}
\end{equation}
In the numerical simulations the winding number of the string will be taken as $n=1$.

Since the Weinberg angle vanishes in this model, the gauge symmetry 
is $SU(2)_L$ while  the $U(1)$ hypercharge decouples and the three
gauge bosons are degenerate. The boson part of the Lagrangian reads
\begin{equation}
\mathcal{L}_{\rm bos}=-\frac{1}{2} {\rm tr}\left(G^{\mu\nu}G_{\mu\nu}\right) 
+\frac{1}{2} {\rm tr} \left(D^{\mu}\Phi \right)^{\dag} D_{\mu}\Phi
- \frac{\lambda}{2} {\rm tr} \left(\Phi^{\dag} \Phi - v^2 \right)^2 \,,
\label{eq:Lboson}
\end{equation}
with the covariant derivative $D_\mu = \partial_\mu - i g W_\mu$ and the 
$SU(2)$ field  strength tensor 
\begin{equation}
G_{\mu\nu}=\partial_\mu\,W_\nu-\partial_\nu W_\mu-ig\left[W_\mu,W_\nu\right]\,.
\label{fieldtensor}
\end{equation}
The boson masses are determined from $g$ and $v$ and the Higgs self-coupling 
$\lambda$ as $m_{\rm W}=gv/\sqrt{2}$ and $m_{\rm H}=2v\,\sqrt{\lambda}$ for the 
gauge and Higgs bosons, respectively. The interaction with the degenerate fermion 
doublet is described by the Lagrangian
\begin{equation}
\mathcal{L}_{\rm fer}=i\overline{\Psi}
\left(P_L \Dslash  + P_R \dslash \right) \Psi
-f\,\overline{\Psi}\left(\Phi P_R+\Phi^\dagger P_L\right)\Psi\,,
\label{eq:Lfermion}
\end{equation}
with the right/left-handed projectors $P_{R,L}=\left(\ID\pm\gamma_5\right)/2$.
Upon spontaneous symmetry breaking, the Yukawa coupling $f$ induces a fermion 
mass~$m=vf$. Assuming a heavy fermion doublet with
the mass of the top quark, the Standard Model suggests the parameters
\begin{equation}
g=0.72\,,\qquad
v=177\,{\rm GeV}\,,\qquad
m_{\rm H}= 140\,{\rm GeV}\,,\qquad
f=0.99\,.
\label{eq:parameters}
\end{equation}
In the numerical search for a stable string, later other values for the Yukawa coupling
will be considered as well.

The classical energy per unit length of the string is determined
by $\mathcal{L}_{\rm bos}$,
\begin{equation}
\frac{E_{\rm cl}}{m^2}=2\pi\int_0^\infty \rho\, d\rho\,\Biggl\{
n^2s_1^2 s_2^2\,\biggl[\frac{2}{g^2}
\left(\frac{f_G^\prime}{\rho}\right)^2
+\frac{f_H^2}{f^2\rho^2}\,\left(1-f_G\right)^2\biggr]
+\frac{f_H^{\prime2}}{f^2}
+\frac{\mu_H^2}{4f^2}\left(1-f_H^2\right)^2\Biggr\}\,,
\label{eq:Ecl}
\end{equation}
where the dimensionless radial integration variable is related to the 
physical radius by $\rho_{\rm phys}=\rho/m$, and we have introduced
the mass ratio $\mu_{H}\equiv m_{\rm H}/m$. 

\subsection{Choice of Gauge and the Dirac equation}

An additional problem arising for string configurations with winding in gauge theories 
is that the string does not induce a well-behaved Born series when expanding in powers 
of the interaction term, $H_{\rm int}$ from Eq.~(\ref{eq:Lfermion}). Even though
the full Hamiltonian is gauge invariant, $H_{\rm int}$ is not and does not
vanish at spatial infinity. This obstacle appears because 
the Dirac Hamiltonian that is obtained by straightforward substitution
of the string background, Eqs.~(\ref{eq:Higgs}) and~(\ref{eq:gauge}),
does not turn into the free Dirac Hamiltonian at $\rho\to\infty$. 
Instead it becomes
$H\to U^\dagger(\varphi) H_{\rm free} U(\varphi)$, where $U(\varphi)$
is a local gauge transformation reflecting the string winding.  It
acts only on the left-handed fermions
\begin{equation}
U(\varphi)=P_L {\rm exp}\left(\imu\hat{\Vek{n}}(\varphi)\cdot\Vek{\tau}\,\xi_1\right)+P_R
\qquad {\rm with} \qquad
\hat{\Vek{n}}(\varphi)=
\begin{pmatrix}
s_2\, {\rm cos}(n\varphi) \cr 
-s_2\, {\rm sin}(n\varphi) \cr c_2
\end{pmatrix}\,.
\label{eq:GT0}
\end{equation}
Unfortunately, making the obvious gauge transformation 
$H \to U(\varphi)HU^\dagger(\varphi)$  does not solve the problem: 
Although it would generate vanishing interactions at infinity, it will also induce a 
$1/\rho^2$ potential at the core of the string, $\rho \to 0$. This choice might still 
yield well-defined phase shifts, though they would likely  be difficult to compute 
numerically, and the conditions underlying  the analyticity of the scattering data 
could be violated by this singular behavior; see, however, Sec.~\ref{ssec:remove}.
As we saw at the end of the previous Section, analyticity is central for numerical 
feasibility of our approach. As a solution, we can define a radially 
extended gauge transformation
\begin{equation}
U(\rho,\varphi)=P_L {\rm exp}\left(\imu\hat{\Vek{n}}\cdot\Vek{\tau}\,\xi(\rho)\right)+P_R\,.
\label{eq:radialGT}
\end{equation}
This transformation fixes the gauge and gives the interaction part
of the Dirac Hamiltonian as
\label{subsec:Gauge}
\begin{align}
H_{\rm int}&=
m\left[\left(f_Hc_\Delta-1\right)
\begin{pmatrix} 1 & 0 \cr 0 &-1\end{pmatrix}
+\imu f_H\,s_\Delta\begin{pmatrix}0 & 1 \cr -1 & 0\end{pmatrix}
\hat{\Vek{n}}\cdot\Vek{\tau}\right]
+\frac{\partial \xi}{2\partial \rho}
\begin{pmatrix}-\Vek{\sigma}\cdot\hat{\Vek{\rho}}
& \Vek{\sigma}\cdot\hat{\Vek{\rho}} \cr
\Vek{\sigma}\cdot\hat{\Vek{\rho}}
& -\Vek{\sigma}\cdot\hat{\Vek{\rho}}\end{pmatrix}\hat{\Vek{n}}\cdot\Vek{\tau}
\cr
&\quad
+\frac{ns_2}{2\rho}\, \begin{pmatrix}-\Vek{\sigma}\cdot\hat{\Vek{\varphi}}
& \Vek{\sigma}\cdot\hat{\Vek{\varphi}} \cr
\Vek{\sigma}\cdot\hat{\Vek{\varphi}}
& -\Vek{\sigma}\cdot\hat{\Vek{\varphi}}\end{pmatrix}
\Big[f_G\,s_\Delta\,I_G(\Delta) +(f_G-1)\,s_\xi\,I_G(-\xi)\Big]\,,
\label{eq:DiracInt} 
\end{align}
where as above we abbreviate $s_\Delta=\sin(\Delta(\rho))$,
$s_\xi=\sin(\xi(\rho))$, etc.
The new gauge function $\xi(\rho)$ appears via the difference 
$\Delta(\rho)\equiv\xi_1-\xi(\rho)$ while the isospin matrices are 
\begin{align}
\hat{\Vek{n}}\cdot\Vek{\tau}&=\begin{pmatrix}
c_2 & s_2 {\rm e}^{\imu n\varphi} \\[2mm] 
s_2 {\rm e}^{-\imu n\varphi} & -c_2 \end{pmatrix}
\quad {\rm and} \quad
I_G(x)&=\begin{pmatrix}
-s_2s_x & (c_2s_x-\imu c_x)\,{\rm e}^{\imu n\varphi} \\[2mm]
(c_2s_x+\imu c_x)\,{\rm e}^{-\imu n\varphi} & s_2s_x\end{pmatrix}\,.
\label{eq:IG}
\end{align}
Imposing the boundary conditions $\xi(0)=0$ and $\xi(\infty)=\xi_1$ for the
new gauge function $\xi(\rho)$ defines a well-behaved scattering problem. The 
specific form of $\xi(\rho)$ is irrelevant (apart from its boundary conditions)
and must not have any influence on the quantum energy, since it merely 
parameterizes a gauge transformation. 

Note that the gauge transformation is single-valued at spatial infinity, 
$U(\infty,\varphi)=U(\infty,\varphi+2\pi)$. In this respect it differs from
the analogous problem of fractional magnetic fluxes in QED. In that case a
similar choice of gauge would not be a remedy; rather the calculation 
of the vacuum polarization energy requires the introduction of a
\emph{return flux} \cite{Graham:2004jb}.  
This approach can also be used for the present calculation,
but it is much more laborious numerically\cite{Weigel:2009wi,Weigel:2010pf}.

We introduce grand-spin type states that couple spin 
and isospin to account for the angular dependence. For fixed angular 
momentum $\ell$ there are four of them,
\begin{equation}
\begin{array}{ll}
\langle \varphi;SI|\ell + +\rangle=
{\rm e}^{i(\ell+n)\varphi}\,
\begin{pmatrix}1 \cr 0 \end{pmatrix}_S\hspace{-2mm}
\otimes\begin{pmatrix}1 \cr 0 \end{pmatrix}_I & \quad
\langle \varphi;SI|\ell + -\rangle =
-\imu\,{\rm e}^{i\ell\varphi}\,
\begin{pmatrix}1 \cr 0 \end{pmatrix}_S\hspace{-2mm}
\otimes\begin{pmatrix}0 \cr 1 \end{pmatrix}_I\cr\cr
\langle \varphi;SI|\ell - +\rangle =
\imu\,{\rm e}^{i(\ell+n+1)\varphi}\,
\begin{pmatrix} 0 \cr 1\end{pmatrix}_S\hspace{-2mm}
\otimes\begin{pmatrix}1 \cr 0 \end{pmatrix}_I & \quad
\langle \varphi;SI|\ell - -\rangle =
{\rm e}^{i(\ell+1)\varphi}\,
\begin{pmatrix}0 \cr 1\end{pmatrix}_S\hspace{-2mm}
\otimes\begin{pmatrix}0 \cr 1 \end{pmatrix}_I\,,
\end{array}
\label{eq:GSstates}
\end{equation}
where $S$ and $I$ refer to the spin and isospin subspaces, respectively.
These grand-spin states serve to construct four-component Dirac spinors in 
coordinate space,
\begin{equation}
\begin{array}{ll}
\langle \rho |++\rangle =
\begin{pmatrix}f_1(\rho)|\ell + +\rangle \cr
g_1(\rho)|\ell - +\rangle \end{pmatrix} & \qquad
\langle \rho |+-\rangle =
\begin{pmatrix}f_2(\rho)|\ell + -\rangle \cr
g_2(\rho)|\ell - -\rangle \end{pmatrix}
\cr \cr
\langle \rho |-+\rangle =
\begin{pmatrix}f_3(\rho)|\ell - +\rangle \cr
g_3(\rho)|\ell + +\rangle \end{pmatrix}& \qquad
\langle \rho |--\rangle =
\begin{pmatrix}f_4(\rho)|\ell - -\rangle \cr
g_4(\rho)|\ell + -\rangle \end{pmatrix}\,,
\end{array}
\label{eq:GSspinors}
\end{equation}
where we have suppressed the angular momentum index of the radial
functions because the Dirac equation is diagonal in this quantum number.
We combine these eight radial functions into two vectors
$\vec{f}=\sum_{i=1}^4f_i(\rho)\hat{\Vek{e}}_i$ and 
$\vec{g}=\sum_{i=1}^4g_i(\rho)\hat{\Vek{e}}_i$
to write the Dirac equation as a set of eight coupled first order 
linear differential equations in the matrix form
\begin{eqnarray}
(E-m)\,\vec{f}&=&V_{uu}\,\vec{f}+\left(D_u+V_{ud}\right)\,\vec{g}\,,\cr
(E+m)\,\vec{g}&=&\left(D_d+V_{du}\right)\,\vec{f}+V_{dd}\,\vec{g}
\label{DiracMatrix}\,.
\end{eqnarray}
The derivative operators are fully contained in the diagonal 
matrices
\begin{eqnarray}
D_u&=&{\rm diag}\left(
\partial_{\rho}+\frac{\ell+n+1}{\rho},
\partial_{\rho}+\frac{\ell+1}{\rho},
-\partial_{\rho}+\frac{\ell+n}{\rho},
-\partial_{\rho}+\frac{\ell}{\rho}\right) \cr\cr
D_d&=&{\rm diag}\left(
-\partial_{\rho}+\frac{\ell+n}{\rho},
-\partial_{\rho}+\frac{\ell}{\rho},
\partial_{\rho}+\frac{\ell+n+1}{\rho},
\partial_{\rho}+\frac{\ell+1}{\rho}\right)\,.
\label{DuDdMatrix}
\end{eqnarray}
We will give the explicit form of the real $4\times4$ matrices $V_i$ in terms of the 
radial functions when we set up the Born series for the scattering data. Here it 
suffices to note that these matrices vanish at spatial infinity, so the asymptotic 
solutions are cylindrical Bessel and Hankel functions. In particular, the Hankel functions
\begin{eqnarray}
\mathcal{H}_u&=&\mbox{diag}\left(
H^{(1)}_{\ell+n}(k\rho),H^{(1)}_{\ell}(k\rho),
H^{(1)}_{\ell+n+1}(k\rho),H^{(1)}_{\ell+1}(k\rho)\right) \cr
\mathcal{H}_d&=&\mbox{diag}\left(
H^{(1)}_{\ell+n+1}(k\rho),H^{(1)}_{\ell+1}(k\rho),
H^{(1)}_{\ell+n}(k\rho),H^{(1)}_{\ell}(k\rho)\right)
\label{Smat2}
\end{eqnarray}
that parameterize the outgoing asymptotic fields with (radial) 
momentum $k$ can be used to set up the scattering problem via the matrix 
generalization
\begin{equation}
\vec{f}\quad \longrightarrow\quad
\mathcal{F} \cdot \mathcal{H}_u
\qquad {\rm and} \qquad
\vec{g}\quad \longrightarrow\quad
\kappa\,\mathcal{G} \cdot \mathcal{H}_d\,,
\label{Smat1}
\end{equation}
where $\kappa=\frac{k}{E+m}=\frac{E-m}{k}$. For simplicity we have not
written out that the matrices $\mathcal{F}$, $\mathcal{G}$ and 
$\mathcal{H}_{u,d}$ are functions of both the radial coordinate~$\rho$ and
the momentum~$k$. The boundary conditions for the $4\times4$ complex
matrices $\mathcal{F}$ and $\mathcal{G}$ are simply 
$\lim_{\rho\to\infty} \mathcal{F} =\ID$ and 
$\lim_{\rho\to\infty} \mathcal{G} =\ID$, so the various columns of the 
above products refer to outgoing waves in different grand spin channels.

The interaction Hamiltonian, Eq.~(\ref{eq:DiracInt}), anti-commutes
with the Dirac  matrix~$\alpha_3$. Hence the spectrum is symmetric and
we only need to consider the case with $E=+\sqrt{k^2+m^2}$ after
analytic continuation to $k=\imu t$. This continuation turns the
Hankel functions into modified Hankel functions  $K_\ell(z)$ and we define
\begin{equation}
Y_u={\rm diag}\, \left(
\frac{K_{\ell+n}(t\rho)}{K_{\ell+n+1}(t\rho)}\,,
\frac{K_{\ell}(t\rho)}{K_{\ell+1}(t\rho)}\,,
-\frac{K_{\ell+n+1}(t\rho)}{K_{\ell+n}(t\rho)}\,,
-\frac{K_{\ell+1}(t\rho)}{K_{\ell}(t\rho)}\right) =
-\left(Y_d\right)^{-1}\,.
\label{defZ12}
\end{equation}
Furthermore we rewrite the kinematic coefficient $\kappa\to z_\kappa$ as
\begin{equation}
z_\kappa=\frac{m+\imu \tau}{t} \,,
\qquad {\rm with}\qquad \tau=\sqrt{t^2-m^2}\,,
\label{kappaAC}
\end{equation}
so that $z_\kappa$ is a pure phase. 

The coupled first-order differential equations take the form
\begin{align}
\partial_\rho \mathcal{F} &=
\left[\mathcal{Q}_{ff}+O_d\right]\cdot\mathcal{F}
+\mathcal{F}\cdot\mathcal{Q}_{ff}^{(r)}
-t\left[\mathcal{Q}_{fg}+C\right]\cdot\mathcal{G}\cdot Y_d
\cr
\partial_\rho \mathcal{G} &=
\left[\mathcal{Q}_{gg}+O_u\right]\cdot\mathcal{G}
+\mathcal{G}\cdot\mathcal{Q}_{gg}^{(r)}
-t\left[\mathcal{Q}_{gf}-C\right]\cdot\mathcal{F}\cdot Y_u\,,
\label{deqGFana}
\end{align}
where the purely kinematical matrices are also straightforwardly
expressed as
\begin{equation}
\mathcal{Q}_{ff}^{(r)}=tC\cdot Y_d(k)-O_d
\qquad {\rm and} \qquad
\mathcal{Q}_{gg}^{(r)}=-tC\cdot Y_u(k)-O_u\,.
\label{ana1}
\end{equation}
The first set of matrices reads
\begin{align}
\mathcal{Q}_{gg}=
\overline{\mathcal{M}}_{gg}
=\begin{pmatrix}
G & P \cr -P & -G^\dagger
\end{pmatrix}  
\qquad {\rm and} \qquad
\mathcal{Q}_{ff}=
\overline{\mathcal{M}}_{ff}
=\begin{pmatrix}
-G^\dagger & P \cr  -P & G
\end{pmatrix} \,.
\label{ana2}
\end{align}
while the matrices involving the energy are
\begin{equation}
\mathcal{Q}_{fg}=\frac{1}{m+\imu\tau}
\begin{pmatrix}
-H & G^\dagger \cr  -G & H
\end{pmatrix}
\qquad {\rm and} \qquad
\mathcal{Q}_{gf}=\frac{1}{m-\imu\tau}
\begin{pmatrix}
H & G \cr  -G^\dagger & -H
\end{pmatrix}\,.
\label{ana3}
\end{equation}
In the above we have conveniently introduced $2\times2$ sub-matrices
\begin{eqnarray}
H &=& \alpha_H\,\begin{pmatrix} 1 & 0 \cr 0 & 1 \end{pmatrix}\,, \qquad\qquad
P = \alpha_P\, \begin{pmatrix} -\imu c_2 & -s_2 \cr s_2 & \imu c_2 \end{pmatrix}=-P^\dagger\,,
\quad {\rm and}
\label{defHGpmP}\\[3mm]
G &=& \alpha_G\, \begin{pmatrix} 
s_2s_\Delta & c_\Delta+\imu c_2s_\Delta \cr
c_\Delta-\imu c_2s_\Delta & -s_2s_\Delta 
\end{pmatrix}
+\alpha_\xi\,\begin{pmatrix} 
-s_2s_\xi & c_\xi-\imu c_2s_\xi \cr
c_\xi+\imu c_2s_\xi & s_2s_\xi \end{pmatrix}
+\alpha_r\, \begin{pmatrix} -\imu c_2 & -s_2 \cr s_2 & \imu c_2 \end{pmatrix}\,.
\nonumber
\end{eqnarray}
The $\alpha$ factors contain the profile functions 
\begin{equation}
\begin{array}{lll}
\alpha_r(\rho)=\frac{1}{2}\,\frac{\partial \xi(\rho)}{\partial\rho} \,,
&\hspace{-1cm}
\alpha_G(\rho)=\frac{ns_2}{2\rho}\,f_G(\rho)\,s_\Delta\,,
&\hspace{1cm}
\alpha_\xi(\rho)=\frac{ns_2}{2\rho}\,\big[f_G(\rho) - 1\big]\,s_\xi\,, \cr\cr
\alpha_H(\rho) = m \,\big[f_H(\rho)\,c_\Delta - 1\big]\, 
&\mbox{\hspace{1cm}and}&\alpha_P(\rho) = m f_H(\rho)\,s_\Delta\,.
\end{array}
\label{alphas}
\end{equation}

We are not yet at the point to compute the Jost function as the logarithm of the determinant 
of $\mathcal{F}$ (or $\mathcal{G}$). These determinants are in
general not real, but rather are complex conjugate to 
each other. This is related to the fact that the (free) spinors explicitly contain
mass factors that differ at $\rho=0$ and $\rho\to\infty$. For real momentum, a typical solution 
in the vicinity of $\rho=0$ looks like~\cite{Bordag:2003at}
\begin{equation}
\begin{pmatrix} f_4 \cr g_4 \end{pmatrix}
\sim \left(\frac{k}{q}\right)^{l}
\begin{pmatrix} \sqrt{E+mc f_H(0)}\, J_l(q\rho) \\[2mm]
\sqrt{E-m cf_H(0)}\, J_{l+1}(q\rho) \end{pmatrix}
\label{regsol}
\end{equation}
with $q=\sqrt{E^2-(m c f_H(0))^2}$,
and similar dependencies hold for the other six radial functions.
These square-root coefficients lead to a definition of the 
Jost function as\cite{Weigel:2009wi}
\begin{equation}
{\rm exp}\left[\nu(t)\right]
=\left(\frac{\tau-\imu m}{\tau-\imu m c f_H(0)}\right)^2 \lim_{\rho\to0}
{\rm det}(\mathcal{F})
=\left(\frac{\tau+\imu m}{\tau+\imu m c f_H(0)}\right)^2 \lim_{\rho\to0}
{\rm det}(\mathcal{G})\,.
\label{defJost}
\end{equation}
The power of two occurs 
because we compute the determinant of a $4\times4$ matrix. Note that 
this redefinition not only cancels the imaginary parts, but also modifies 
the real part. Furthermore, it cancels the logarithmic singularity in 
${\rm ln}\lim_{\rho\to0} {\rm det}(\mathcal{F})$ observed numerically at 
$t\sim m$. Since $f_H$ is part of the interaction, this correction factor 
also undergoes expansion in the framework of the Born series. The Born series 
for ${\rm det}(\mathcal{F})$ is constructed by iterating the differential 
equations of Eq.~(\ref{deqGFana}) in $\overline{\mathcal{M}}_i$; see
Ref.\cite{Graham:2011fw} for more details.

\subsection{Numerical set-up}

Thus far we have addressed the technical obstacles related to the singular structure
of the string at the origin. Despite the additional simplification due
to the use of a fake boson 
subtraction as in Eq.~(\ref{eq:fb1}) for the subleading logarithmic divergences, the numerical 
computation is still expensive. The scattering data are extracted from a multi-channel problem 
and, for the final result to be reliable, several hundred partial wave channels must be 
included. Furthermore, channels that contain orbital angular momentum $\ell=0$ require 
disentangling a constant from a logarithm for the regular and irregular solutions
when $\rho=\rho_{\rm min}\to0$. This is only possible by extrapolating
a fit of the form
\begin{equation}
{\rm det}(\mathcal{F})=a_0 +\frac{a_1}{{\rm ln}(\rho_{\rm min})}
+\frac{a_2}{{\rm ln}^2(\rho_{\rm min})}\ldots\, \longrightarrow\, a_0\,,
\label{nuextra}
\end{equation}
for the Jost function in these channels. These numerical efforts restrict
the number of variational parameters that can be used to characterize the
profile functions. In addition to $\xi_1$ and $\xi_2$, we introduce three 
scale parameters $w_H$, $w_W$, and $w_\xi$ via the ans\"atze
\begin{equation}
f_H(\rho)=1-{\rm e}^{-\rho/w_H}\,,\qquad
f_G(\rho)=1-{\rm e}^{-\left(\rho/w_G\right)^2}\,,\qquad
\xi(\rho)=\xi_1\left[1-
{\rm e}^{-\left(\rho/w_\xi\right)^2}\right]\,.
\label{eq:Ansaetze}
\end{equation}
The scale $w_\xi$ parameterizes the shape of the gauge profile, which
should not be observable
as discussed above. The radial dependencies of the profiles are chosen to keep $E_{\rm cl}$ regular.
In Ref.~\cite{Graham:2011fw} other parameterizations for $f_G(\rho)$ have also been considered.
No substantial differences for the VPE were observed.

In what follows we write the VPE as the sum
\begin{equation}
\Delta E=E_{\rm scat}+E_{\rm FD}\,,
\label{eq:VPE_sum}
\end{equation}
where $E_{\rm scat}$ is the momentum integral as in Eq.~(\ref{eq:fb1}) and $E_{\rm FD}$
is the combination of all Feynman diagrams with the counterterms. 

\subsection{Gauge invariance}

We check gauge invariance by varying the shape of the gauge profile,
$\xi(\rho)$. Typical results are shown in Tab.~\ref{tab:Res1}.
\begin{table}[pb]
\tbl{Numerical results for the fermion vacuum polarization energy $\Delta E$ 
adapted from Ref.\cite{Weigel:2010pf} for $w_G=w_H=2.0$ and $\xi_1=0.4\pi$. 
The second row shows the contribution from the renormalized Feynman diagrams 
in the minimal subtraction scheme (including the fake boson piece) while
$E_\delta$ denotes the scattering momentum integral. The last two lines originate
from an adjusted large $t$ treatment of the integral for $E_{\rm scat}$, see text.}
{\begin{tabular}{l|cccc}
~~$w_\xi$~~ & 1.0 & 2.0 & 3.0 & 4.0\cr
\hline
~~$E_{\rm FD}$~~ &~~ -0.0623~~ &~~ -0.0320~~ &~~ -0.0264~~ &~~ -0.0222 \cr
~~$E_{\rm scat},\,\alpha=2.0$~ &~~ 0.1606~~ &~~ 0.1294 &~~ 0.1235~~ &~~ 0.1193\cr
\hline
~~$\Delta E$~~   &~~ 0.0983~~ &~~ 0.0974~~ &~~ 0.0971~~ &~~ 0.0971 \cr
\hline\hline
~~$E_{\rm scat},\,\alpha=2.4$~ &~~ 0.1588~~ &~~ 0.1280~~ &~~ 0.1222~~ &~~ 0.1188\cr
\hline
~~$\Delta E$~~   &~~ 0.0964~~ &~~ 0.0960~~ &~~ 0.0958~~ &~~ 0.0959
\end{tabular}}
\label{tab:Res1}
\end{table}
As expected, the individual contributions to $\Delta E$ depend strongly on
$w_\xi$. However, these changes cancel out almost completely. Numerically
the most cumbersome part of the calculation is $E_{\rm scat}$. From various
numerical considerations (change of extrapolation scheme for the partial wave sum,
modification of the momentum integration grid, etc.), its numerical accuracy is
estimated to be at the 1\% level. An example for the accuracy test is presented
in the last two lines of Tab.~\ref{tab:Res1}. The large $t$ contribution to
the integral in Eq.~(\ref{eq:inter2}) is computed by fitting an inverse power 
to the integrand: $\frac{1}{t^\alpha}$. The large mass expansion of the Feynman
diagrams yields $\alpha=2.0$, while a numerical fit has a slightly larger value 
$\alpha\approx2.4$, most likely because subleading powers are not fully negligible. 
Within that range of numerical uncertainty of $E_{\rm scat}$, $\Delta E$
is indeed independent of $w_\xi$.

\subsection{Results for on-shell renormalization}
\label{ssec:onshell}

The results in Tab.~\ref{tab:Res1} were obtained in the $\overline{\rm MS}$ 
renormalization scheme, which essentially omits finite terms 
introduced with renormalizing the 
Feynman diagrams. Any other scheme merely differs by manifestly gauge
invariant, finite counterterms. To obtain physically meaningful results, we need to impose renormalization 
conditions that correspond to a particle interpretation. To be specific, we consider 
the so-called {\it on-shell} scheme, in which the coefficients of the four allowed 
counterterms are determined such that
\begin{itemize}
\item[$\bullet$]
the tadpole graph vanishes
\item[$\bullet$]
the Higgs mass remains unchanged
\item[$\bullet$]
the normalization of Higgs particle remains unchanged
\item[$\bullet$]
and the normalization of vector meson remains unchanged
\end{itemize}
in the presence of fermionic quantum corrections. Explicit
expressions for the corresponding counterterms are listed in
Ref.\cite{Graham:2011fw}, and
a similar calculation is sketched in Sec.~\ref{ssec:topcharge}. 
Note that the vector meson mass~$M_W$ is not fixed by these conditions
and thus will be a prediction that includes quantum corrections. Hence we
tune the gauge coupling to reproduce the physical value
$M_W\approx90\,{\rm GeV}$.

\subsubsection{$\xi_2=\frac{\pi}{2}$}
\label{sss:pihalf}

In a first step we consider the particular case $\xi_2=\frac{\pi}{2}$, for which the 
Dirac Hamiltonian not only is Hermitian but also yields real matrix elements in
Eq.~(\ref{defHGpmP}).

\begin{figure}[tp]
\centerline{~\hspace{-0.5cm}
\includegraphics[width=6.6cm,height=3.7cm]{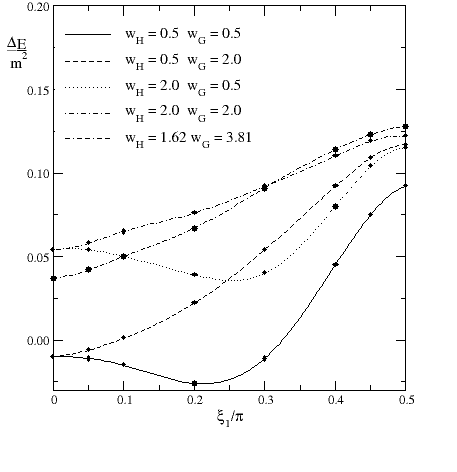}~~
\includegraphics[width=6.2cm,height=4.1cm]{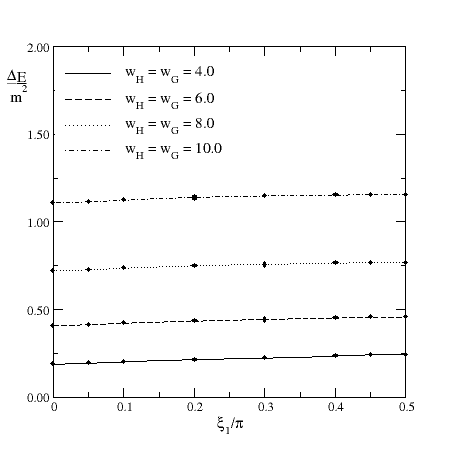}}
\caption{\label{fig:CSVPE}Vacuum polarization energy as a function of the angle $\xi_1$
for different values of the width parameters $w_H$ and $w_G$ in the on-shell 
renormalization scheme. The physically motivated model parameters, Eq.~(\ref{eq:parameters}), 
are used. The dots refer to actual computations, while the lines stem from a cubic spline. 
These results do not include the combinatoric color factor $N$.}
\end{figure}

In Fig.~\ref{fig:CSVPE} we show results for the vacuum polarization energy per unit length.
The wider the background fields, the weaker the dependence on the angle $\xi_1$. The vacuum 
polarization per unit length is quite small. Even for large widths, it does not exceed a 
fraction of the fermion mass squared. With the exception of very small widths, the vacuum 
polarization turns out to be positive. Hence there is no indication that the fermion vacuum 
polarization energy alone can stabilize cosmic strings, since the classical energy is larger 
by orders of  magnitude unless the  coupling constants are $f,g\sim{\mathcal O}(10)$, which 
would bring the dominating Fourier components of the profiles into the vicinity of the Landau ghost pole. 
Hence any such binding would be obscured by the existence of the Landau ghost, which arises when 
including quantum corrections in a manner that does not reflect asymptotic freedom. Here it is 
due to the omission of quantum corrections from fluctuating gauge boson fields. The estimate for 
the Landau ghost contribution discussed in Ref.~\cite{Graham:2011fw} suggests that the issue can 
be safely ignored for $f,g\lesssim5$.

\subsubsection{Isospin invariance}

Alternatively, the field configuration in Eqs.~(\ref{eq:Higgs}) and~(\ref{eq:gauge}) can be 
written as
\begin{align}
\begin{pmatrix} \phi_+(\rho,\varphi) \\[1mm]
\phi_0(\rho,\varphi) \end{pmatrix}&=
 f_H(\rho) U(\xi_1,\xi_2,\varphi)
\begin{pmatrix}0 \\[1mm] v \end{pmatrix}
\qquad {\rm and} \cr
\Vek{W}(\rho,\varphi)&= \frac{1}{g}\,
\frac{\hat{\Vek{\varphi}}}{\rho} f_G(\rho)\,
U(\xi_1,\xi_2,\varphi)\,\partial_\varphi
U^\dagger(\xi_1,\xi_2,\varphi)\,.
\label{eq:profiles}\end{align}
This formulation has introduced the $SU(2)$ matrix
\begin{equation}
U(\xi_1,\xi_2,\varphi)=n_0\ID-i\Vek{n}\cdot\Vek{\tau}
\quad{\rm with}\quad
\hat{\Vek{n}}_4(\xi_1,\xi_2,\varphi)=\begin{pmatrix}
{\rm sin}\xi_1\,{\rm sin}\xi_2\, {\rm cos}\varphi\cr
{\rm cos}\xi_1 \cr {\rm sin}\xi_1 \,{\rm cos}\xi_2\cr
{\rm sin}\xi_1 \,{\rm sin}\xi_2\, {\rm sin}\varphi
\end{pmatrix}\,.
\label{eq:nhat}
\end{equation}
A global rotation within the plane of the second and third component by the 
angle~$\alpha$ with ${\rm tan}\alpha=s_1c_2/c_1$ transforms the four-component
unit vector $\hat{\Vek{n}}_4$ into
\begin{equation}
\widetilde{\Vek{n}}_4(\xi_1,\xi_2,\varphi)=\begin{pmatrix}
s_1s_2 {\rm cos}\varphi\cr \sqrt{1-s_1^2s_2^2} \cr
0 \cr s_1s_2 {\rm sin}\varphi
\end{pmatrix}\,.
\label{eq:nhat1}
\end{equation}
Hence observables (which are, by definition, gauge invariant) will not
depend on the two angles $\xi_1$ and $\xi_2$ individually but only on the
product $s_1s_2$. Said another way, all observables must remain invariant 
along paths of constant $s_1s_2$ in isospin space~\cite{Klinkhamer:1997hw}.
This invariance is not manifest for our calculation of the VPE. For example,
the local gauge transformation, Eq.~(\ref{eq:radialGT}), does not exhibit 
that invariance. Neither the individual Feynman diagrams that are 
added back for the subtraction of Born terms, nor the fake boson method 
are subject to that symmetry. We therefore have to verify that invariance
from the numerical simulation of the full VPE.

\begin{table}[htbp]
\tbl{\label{tab:zz} Contributions to Eq.~(\ref{eq:VPE_sum}) and their
variation with the isospin angles. In all cases we have $s_1s_2\approx0.29389$.
The width parameters of the boson profiles are $w_G=w_H=3.5$. The results were
obtained with various values for the widths of the gauge and fake boson profiles.}
{\begin{tabular}{c|c|c|c||c||c}
$\xi_1/\pi$ &$\xi_2/\pi$ & $E_{\rm scat}$ &  $E_{\rm FD}$ & $\Delta E$
& $|E_{\rm scat}|+|E_{\rm FD}|$\cr
\hline
0.1 & 0.4 & 0.1504 & 0.0014 &~~0.1518~~~& 0.1518\cr
0.4 & 0.1 & 0.1702 & -0.0180 &~~0.1521~~~& 0.1882 \cr
0.3 & 0.11834 & 0.1496 & 0.0021 &~~0.1517~~~& 0.1517 \cr
0.2 & 1/6 & 0.1639 & -0.0117 &~~0.1522~~~& 0.1758
\end{tabular}}
\end{table}
From the sample calculations listed in Tab.~\ref{tab:zz} we see that the
variation in $\Delta E$ along a path with constant $s_1s_2$ is as small as 
a quarter of a percent, while the absolute values of the contributions 
vary on the order of 20\%.

\begin{figure}
\centerline{
\includegraphics[width=6.5cm,height=4.5cm]{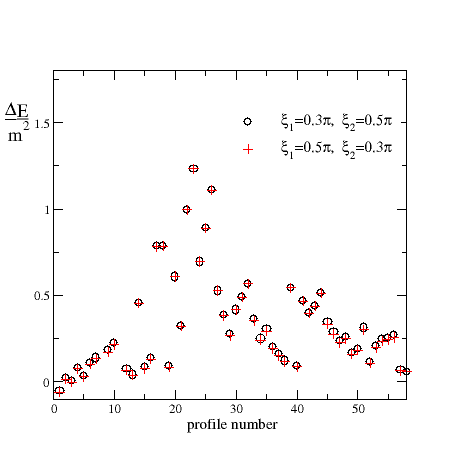}~~~
\includegraphics[width=6.5cm,height=4.5cm]{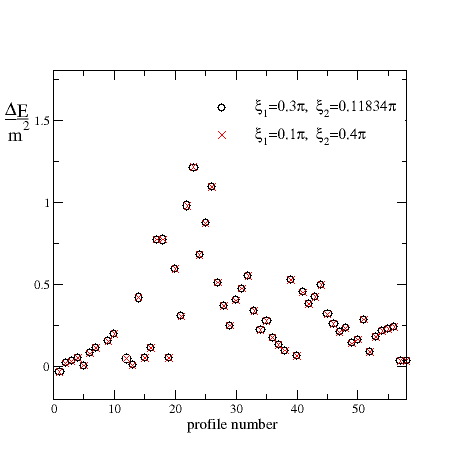}}
\caption{\label{fig:x1x2}The vacuum polarization energy for different background profiles 
for two sets of angles. Each set has the same product $s_1s_2$. The profile counting goes 
by different values of the width parameters $w_G$ and $w_H$~\cite{Weigel:2016ncb}.}
\end{figure}

The simplest check of two configurations with the same $s_1s_2$ is
just to swap the two angles. 
Results of that operation are shown in the left panel of Fig.~\ref{fig:x1x2} 
for several dozen profile functions characterized by different $w_G$ and $w_H$.
Of course, swapping the two angles would not change the VPE if it were a function
of their sum. We therefore also present in the right panel of Fig.~\ref{fig:x1x2} the 
results from an alternative set of angles with identical $s_1s_2$. Ref.~\cite{Weigel:2016ncb} 
verifies the isospin invariance for more sets of angles. With the confirmation of 
this invariance, we conclude that $\xi_2$ is a redundant variational parameter and 
we may without loss of generality
simplify to the case of $\xi_2=\frac{\pi}{2}$, for which the Dirac Hamiltonian 
has real matrix elements.

\subsubsection{Stable charged strings}

Although the VPE alone does not stabilize classical string configurations, strings with 
fermion charge can potentially be stabilized by having lower energy than the same number 
of free fermions.  Wide strings in particular generate many fermion bound states. 
Their energy eigenvalues are of the same order in the semi-classical $\hbar$ expansion as 
the VPE, however, so the inclusion of these levels ultimately requires
consideration of the VPE as well. 

We want to compute the binding energy per unit length for a prescribed charge per unit 
length $Q$. Let $0<\omega_i\le m$ be a bound state eigenvalue of the Dirac Hamiltonian 
whose interaction part is given in Eq.~(\ref{eq:DiracInt}). Then a state has energy 
$\left[\omega_i^2+p^2\right]^{\sfrac{1}{2}}$, where $p$ is the conserved momentum 
along the symmetry axis. To count the populated states, we introduce a chemical 
potential $\mu$ such that ${\rm min}\{\omega_i\}\le\mu\le m$.  States with
$[\omega_i^2+p^2]^{\sfrac{1}{2}}<\mu$ are filled while states with 
$[\omega_i^2+p^2]^{\sfrac{1}{2}}>\mu$ remain empty. For each bound state, this
defines the Fermi momentum $P_i(\mu)=
[\mu^2-\omega_i^2]^{\sfrac{1}{2}}$, and the charge (per unit length) is
\begin{equation}
Q(\mu)=\sum_{\omega_i\le\mu}\int_{-P_i(\mu)}^{P_i(\mu)} \frac{dp}{2\pi}=
\frac{1}{\pi}\sum_{\omega_i\le\mu} P_i(\mu)\,.
\label{eq:charge}
\end{equation}
The sum runs over all bound states available for a given chemical potential. This 
relation can be inverted to give $\mu=\mu(Q)$: For prescribed~$Q$, we increase~$\mu$ 
from ${\rm min}\{\omega_i\}$ until the right-hand-side of Eq.~(\ref{eq:charge})
matches. The same number of non-interacting fermions has energy of at
least $Qm$, giving the binding energy\footnote{Here we define the
binding energy such that a negative value indicates binding.}
per unit length
\begin{equation}
E_{\rm b}(Q)=E_{\rm cl}+N\Delta E+
\frac{N}{\pi}\sum_i\int_0^{P_i(\mu(Q))}dp \,
\left[\sqrt{\omega_i^2+p^2}-m\right]\,.
\label{eq:ebind}
\end{equation}
For a prescribed charge $Q$, we find an upper bound on $E_{\rm b}(Q)$ by scanning several 
hundred configurations, parameterized by different values of $w_H$, $w_G$ and~$\xi_1$.

This model is similar to the Standard Model of particle physics and we thus adopt the 
parameters from Eq.~(\ref{eq:parameters}) together with $N=3$ (for color). Eventually we 
will allow for a heavy (fourth) generation of fermions and thus vary the Yukawa coupling $f$.

\begin{figure}
\centerline{
\includegraphics[width=6.3cm,height=4.5cm]{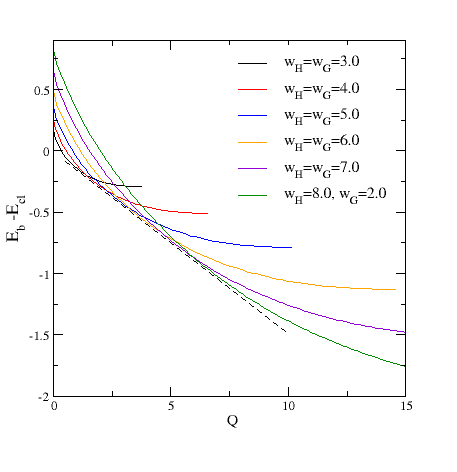}~~~
\includegraphics[width=6.8cm,height=4.7cm]{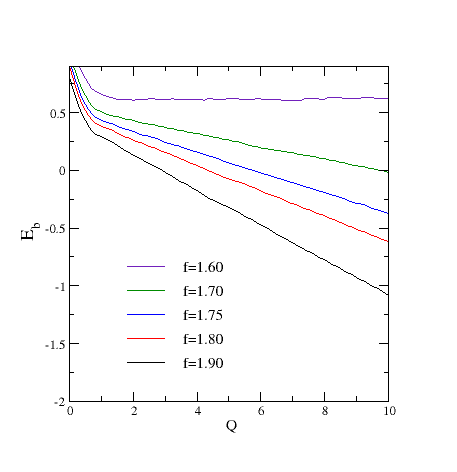}}
\caption{\label{fig:charge}For the model parameters of Eq.~(\ref{eq:parameters})
the left panel shows the fermion energy when $Q$ levels are occupied for selected variational
parameters at $\xi_1=0.4\pi$. The optimal value follows a straight line. The right panel gives 
the optimal binding energy as function of the charge for several Yukawa constants. All other model 
parameters are from Eq.~(\ref{eq:parameters}).}
\end{figure}

The left panel of Fig.~\ref{fig:charge} shows the fermion contribution to the energy $E_{\rm b}$ 
as a function of the charge. As mentioned in Sec.~\ref{sss:pihalf}, for wide profiles there is 
only very mild dependence on $\xi_1$ and the specific value is not essential. Those graphs 
terminate at the charge when all bound states are occupied. The energy $E_{\rm b}$ is smaller 
the wider the string configuration because the number of available bound states increases with 
the width, in particular with the width of the scalar component of the Higgs field.

The right panel of Fig.~\ref{fig:charge} gives the final result for the binding energy. When measured
in units of the fermion mass $m^2$, the fermion contribution does not scale with the Yukawa coupling $f$
while the Higgs contribution to the classical energy, Eq.~(\ref{eq:Ecl}), scales like $\frac{1}{f^2}$. 
Hence for a large enough coupling constant, $E_{\rm b}$ will dominate. From that figure we find that
the critical value above which binding is observed is about 
$f_{\rm c}\approx1.7$, corresponding
to a fermion with mass of approximately $300{\rm MeV}$. The larger the Yukawa coupling, the smaller
the charge of the bound configuration.

We have not included the interaction among the fermions that build the charge. This is similar to the
description of atomic levels from the hydrogen spectrum. Also, the full back-reaction on the meson profiles
is not included. However, we stress that since we are using a the variational approach, such a modification 
can only lower the total energy.

\section{Vortices}
\label{sec:vortices}

In this Section we compute the VPE of Abrikosov-Nielsen-Olesen (ANO) vortices 
\cite{Abrikosov:1956sx,ABRIKOSOV1957199,Nielsen:1973cs} for different
topological charges. There are multiple applications of such vortices in physics: In condensed 
matter physics, the dependence of the energy on the topological charge is essential for 
distinguishing superconductors of type I, where the energy increases
weaker than linearly with charge and multiple vortices coalesce, from
type II, where energy increases stronger than linearly and single, isolated 
vortices emerge. In particle physics applications the topological charge is often identified as 
the particle number; therefore these studies provide important insight for binding energies beyond 
the classical level.

These vortices consist of a scalar Higgs field with spontaneous symmetry breaking and an Abelian
gauge field. For simplicity we consider the BPS case with equal masses, which classically represents 
the transition between type I and type II superconductors.

\subsection{Classical Fields}

Classical vortices are constructed from the Lagrangian of scalar electrodynamics
with spontaneous symmetry breaking
\begin{equation}
\mathcal{L}=-\frac{1}{4}F_{\mu\nu}F^{\mu\nu}+|D_\mu\Phi|^2
-\frac{\lambda}{4}\left(|\Phi|^2-v^2\right)^2\,,
\label{eq:lag1}\end{equation}
where $F_{\mu\nu}=\partial_\mu A_\mu-\partial_\nu A_\mu$ is the field strength
tensor and $D_\mu\Phi=\left(\partial_\mu-\imu e A_\mu\right)\Phi$ is
the covariant derivative of the complex Higgs field.

We will need to consider fluctuations about vortex backgrounds that approach free cylindrical
waves as $\rho\to\infty$. This requires the so-called singular gauge, which is characterized 
by two profile functions $h$ and $g$ in
\begin{equation}
\Phi_S=vh(\rho) 
\qquad {\rm and}\qquad 
\Vek{A}_S=nv\hat{\Vek{\varphi}}\,\frac{g(\rho)}{\rho}
\label{eq:string1}\end{equation}
where $\rho=evr$ is dimensionless while $r$ is the physical distance
from the vortex, and furthermore $\hat{\Vek{\varphi}}$ is the azimuthal unit vector for the vortex axis. The 
temporal and longitudinal components of the gauge field vanish classically, $A_S^0=A_S^3=0$.
The winding number $n$ is the essential topological quantity. In the BPS case with
$\lambda=2e^2$, the energy functional is minimized when the profile functions
obey the first-order differential equations
\begin{equation}
g^\prime=\frac{\rho}{n}(h^2-1)
\qquad {\rm and}\qquad
h^\prime=\frac{n}{\rho}gh\,,
\label{eq:cldeqBPS}\end{equation}
with the singular gauge boundary conditions
\begin{equation}
h(0)= 1 - g(0)=0 \hbox{~~~~and~~~~}
\lim_{\rho\to\infty}h(\rho) =
1-\lim_{\rho\to\infty}g(\rho)=1\,.
\label{eq:bcANO}\end{equation}
The resulting energy per unit length is linear in the winding number,
$E_{\rm cl}=2\pi n v^2$. The differential
equations~(\ref{eq:cldeqBPS}) can be solved numerically, but for
later use in the scattering problem an approximate expression in terms
of elementary functions is very helpful. It turns out that for $1\le
n\le4$ the correlation coefficients for the fit
\begin{align*}
h(\rho)&=\alpha_2\tanh^n(\alpha_1\rho)+[1-\alpha_2]\tanh^n(\alpha_0\rho)\cr
g(\rho)&=\beta_1\rho\,\frac{1-\tanh^2(\beta_2\rho)}{\tanh(\beta_1\rho)}
\end{align*}
with the fit parameters $\alpha_i$ and $\beta_i$ listed in Tab.~\ref{tab:fit} deviate from 
unity by~$10^{-4}$ or less from the numerical solutions to Eq.~(\ref{eq:cldeqBPS}).
\begin{table}[ht]
\tbl{\label{tab:fit}Fit parameters for vortex profiles.}
{\begin{tabular}{c|ccc|cc}
$n$ & $\alpha_0$ & $\alpha_1$ & $\alpha_2$ & $\beta_1$ & $\beta_2$\cr
\hline
1 & 0.8980 & 0.6621 & 0.1890 & 0.5361 & 0.7689 \\
2 & 0.9072 & 0.8288 & 2.6479 & 1.0949 & 0.8042 \\
3 & 0.8290 & 0.7882 & 5.1953 & 1.1328 & 0.7425 \\
4 & 0.7755 & 0.7350 & 5.2009 & 1.1034 & 0.6853
\end{tabular}}
\end{table}

\subsection{Quantum theory}

To quantize the theory, we introduce fluctuations about the vortex via
\begin{equation}
\Phi=\Phi_S+\eta \qquad {\rm and}\qquad 
A^\mu=A_S^\mu+a^\mu
\label{eq:fluct}\end{equation}
and extract the harmonic terms in the fluctuations $\eta$ and $a^\mu$. Their
gauge is fixed by adding an $R_\xi$ type Lagrangian that cancels the 
$\eta\partial_\mu a^\mu$ and $\eta^\ast\partial_\mu a^\mu$ terms,
\begin{equation}
\mathcal{L}_{\rm gf}=-\frac{1}{2}G^2=-\frac{1}{2}\left[\partial_\mu a^\mu
+\imu e\left(\Phi_S\eta^\ast-\Phi_S^\ast\eta\right)\right]^2\,.
\label{eq:laggf}\end{equation}
We still have to account for the ghost contribution to the VPE associated with this gauge fixing.
The infinitesimal gauge transformations read
\begin{equation}
A^\mu\to A^\mu+\partial^\mu\chi\,, \qquad 
\Phi_S+\eta\to \Phi_S+\eta+\imu e\chi(\Phi_S+\eta)
\end{equation}
so that $a^\mu\to a^\mu+\partial^\mu\chi$ and $\eta\to\eta +\imu e\chi(\Phi_S+\eta)$. Then
\begin{equation}
\frac{\partial G}{\partial \chi}\Big|_{\chi=0}
=\partial_\mu\partial^\mu+e^2\left(2|\Phi_S|^2+\Phi_S\eta^\ast+\Phi_S^\ast\eta\right)
\label{eq:fp1}\end{equation}
induces the ghost Lagrangian \cite{Lee:1994pm,Rebhan:2004vu}
\begin{equation}
\mathcal{L}_{\rm gh}=\overline{c}\left(\partial_\mu\partial^\mu+2e^2|\Phi_S|^2\right)c +
\mbox{non-harmonic terms}\,.
\label{eq:fp2}\end{equation}
The corresponding VPE is that of a Klein-Gordon field with mass $\sqrt{2} ev$ 
in the background potential $2v^2(h^2-1)$, which must be multiplied by 
a factor of negative two, corresponding to a complex scalar ghost
field, and combined with the VPE obtained for the gauge and scalar fields.
Since $D_0\Phi_S=0$ and $D_3\Phi_S=0$, the temporal and longitudinal
components $a^0$ and $a^3$ fully decouple, contributing 
$$
-\frac{1}{2}\left[\partial_\mu a_0\partial^\mu a^0+\partial_\mu a_3\partial^\mu a^3\right]
+|\Phi_S|^2\left[a_0a^0+a_3a^3\right]
$$
to the Lagrangian. These fluctuations are both subject solely to the background potential 
$2(|\Phi_S|^2-v^2)$, which is exactly the same as that of the ghosts. As a result, 
the non-transverse and ghost contributions to the VPE cancel each other. Of course,
this just reflects the fact that the free electromagnetic field only
has two physical degrees of freedom.
Thus we end up with the truncated Lagrangian for the relevant quantum fluctuations,
\begin{align}
\mathcal{L}^{(2)}
&=\frac{1}{2}\sum_{n=1,2}\left(\partial_\mu a_n\right)\left(\partial^\mu a_n\right)
-e^2|\Phi_S|^2\sum_{n=1,2}a_n^2\cr
&\hspace{0.5cm}
+|\dot{\eta}|^2-|\partial_3\eta|^2+\sum_{n=1,2}(D_n\eta)^\ast(D^n\eta)
-e^2\left[3|\Phi_S|^2-v^2\right]|\eta|^2\cr
&\hspace{0.5cm}
+2\imu e\sum_{n=1,2}a_n\left[\eta^\ast \left(D^n\Phi_S\right)-\eta \left(D^n\Phi_S\right)^\ast\right]\,.
\label{eq:fluc}\end{align}
Essentially we have simplified the theory to that of four real scalar fields:
$a_1$, $a_2$, ${\sf Re}(\eta)$ and ${\sf Im}(\eta)$.

To formulate the scattering problem, we employ a partial wave decomposition using the 
complex combinations in units with $ev=1$ (resulting in mass parameters $M_H=M_A=\sqrt{2}$),
\begin{align}
a^1+\imu a^2=\sqrt{2}\imu{\rm e}^{-\imu\omega t}\sum_\ell a_\ell(\rho){\rm e}^{\imu \ell \varphi}
\quad {\rm and}\quad
\eta={\rm e}^{-\imu\omega t}\sum_\ell\eta_\ell(\rho){\rm e}^{\imu \ell \varphi}\,,
\label{eq:partialwave}
\end{align}
leading to a $4\times4$ scattering problem for the radial 
functions. For profile functions obeying Eq.~(\ref{eq:cldeqBPS}),
this problem decouples into two $2\times2$ systems, with the one for $a^\ast$ and $\eta^\ast$
being the same as that of $a$ and $\eta$. Hence it suffices to compute the VPE of the
latter and double it. The scattering problem is set up in terms of the Jost solution
$\mathcal{F}_\ell$ by introducing
\begin{equation}
\begin{pmatrix}\eta^{(1)}_\ell & \eta^{(2)}_\ell \cr 
a^{(1)}_{\ell+1} & a^{(2)}_{\ell+1} \end{pmatrix}=\mathcal{F}_\ell\cdot\mathcal{H}_\ell
\quad {\rm where}\quad
\mathcal{H}_\ell=\begin{pmatrix}H^{(1)}_{\ell}(q\rho)& 0
\cr 0& H^{(1)}_{\ell+1}(q\rho)\end{pmatrix}\,.
\label{eq:param}\end{equation}
The superscripts on the left-hand side refer to the two possible scattering 
channels when imposing the boundary condition $\lim_{\rho\to\infty}\mathcal{F}_\ell=\ID$.
The Hankel functions $H_\ell^{(1)}$ parameterize outgoing cylindrical waves.
In matrix form, the scattering differential equation in terms of the 
dimensionless imaginary momentum $t=\imu \sqrt{\omega^2-2}$ reads
\begin{equation}
\frac{\partial^2}{\partial\rho^2}\mathcal{F}_\ell
=-\frac{\partial}{\partial\rho}\mathcal{F}_\ell
-2\left(\frac{\partial}{\partial\rho}\mathcal{F}_\ell\right)\cdot\mathcal{Z}_\ell
+\frac{1}{\rho^2}\left[\mathcal{L}_\ell,\mathcal{F}_\ell\right]
+\mathcal{V}_\ell\cdot\mathcal{F}_\ell\,,
\label{eq:jostdeq}\end{equation}
where angular momenta enter via the derivative matrix for the analytically continued 
Hankel functions
\begin{equation}
\mathcal{Z}_\ell=\begin{pmatrix}
\frac{|l|}{\rho}-t\,\frac{K_{|l|+1}(t\rho)}{K_{|l|}(t\rho)} & 0 \cr
0 & \frac{|l+1|}{\rho}-t\,\frac{K_{|l+1|+1}(t\rho)}{K_{|l+1|}(t\rho)}
\end{pmatrix}
\quad {\rm and}\quad
\mathcal{L}_\ell=\begin{pmatrix}\ell^2 & 0 \cr 
0 & (\ell+1)^2 \end{pmatrix}\,.
\label{eq:angmom}\end{equation}
The potential matrix is
\begin{equation}
\mathcal{V}_\ell=\begin{pmatrix}
3(h^2(\rho)-1)+\frac{n^2g^2(\rho)-2n\ell g(\rho)}{\rho^2}
& \sqrt{2}d(\rho)\cr
\sqrt{2}d(\rho)& 2(h^2(\rho)-1) \end{pmatrix}\,,
\label{eq:potmat}\end{equation}
with $d(\rho)=\frac{dh(\rho)}{d\rho}+\frac{n}{\rho}h(\rho)g(\rho)$.
We then use Eq.~(\ref{eq:jostdeq}) to compute the Jost function, which
is given by $\nu_\ell(t)=\lim_{\rho\to0}{\rm ln}
\,{\rm det}\left[\mathcal{F}_\ell\right]$.

\subsection{Removal of gauge-variant divergence}
\label{ssec:remove}

\begin{figure}[ht]
\includegraphics[width=13cm,height=2.5cm]{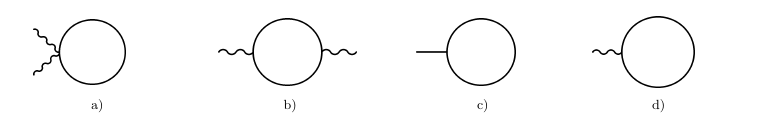}
\caption{Quadratically divergent one-loop diagrams with a Higgs field in the loop. 
External lines represent (Fourier transforms of) the classical Higgs (straight) 
and gauge (curly) fields.}
\label{fig:Qdiv}
\end{figure}

\begin{figure}[ht]
\includegraphics[width=13cm,height=2.5cm]{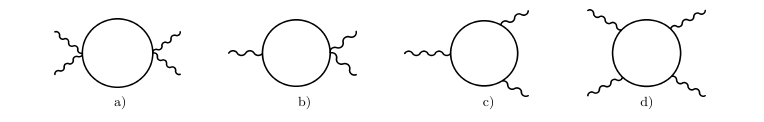}
\caption{Logarithmically divergent one-loop Higgs diagrams with
external photon lines only.}
\label{fig:LdivA}
\end{figure}

\begin{figure}[ht]
\includegraphics[width=13cm,height=2.5cm]{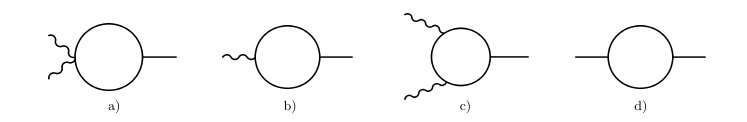}
\caption{Logarithmically divergent one-loop Higgs diagrams with
at least one insertion of the Higgs potential $V_H=3(h^2-1)$.}
\label{fig:LdivB}
\end{figure}

The Higgs-Higgs component of the potential matrix in Eq.~(\ref{eq:potmat}) diverges like
$\frac{1}{\rho^2}$ as $\rho\to0$. This singular behavior has neither well-defined Born 
nor Feynman series. Hence we need to develop an alternative method to handle the associated 
ultra-violet divergences. To study these divergences in more detail, we display all divergent 
one-loop diagrams arising from Higgs fluctuations in Fig.~\ref{fig:Qdiv}, \ref{fig:LdivA}
and \ref{fig:LdivB}.
Fig.~\ref{fig:Qdiv} shows the Feynman diagrams that superficially are quadratically 
divergent. Due to gauge invariance, the quadratic divergences of \ref{fig:Qdiv}a) and \ref{fig:Qdiv}b) cancel. 
Diagram \ref{fig:Qdiv}d) is a total derivative and vanishes (with a translationally invariant 
regularization). Indeed all diagrams with an odd number of gauge field insertions are 
finite because of Lorentz invariance and the fact that $\partial_\mu A_S^\mu=0$. Hence
the only remaining quadratic divergence is the tadpole graph with a single insertion of 
$V_H$, as shown in \ref{fig:Qdiv}c). This diagram  is local, meaning it is
independent of the incoming momentum and thus
proportional to $\int d^2x V_H$, so it can be fully removed from the VPE by an appropriate 
{\it no-tadpole} renormalization condition. Again by gauge invariance, the logarithmic 
divergences in \ref{fig:LdivA}a) and \ref{fig:LdivA}d) cancel, as do those of~\ref{fig:LdivB}a) 
and~\ref{fig:LdivB}c). Thus all we need to consider are the divergences associated with the 
diagrams of Fig.~\ref{fig:Qdiv}a)-c) and~\ref{fig:LdivB}d). The treatment of \ref{fig:Qdiv}c) 
and \ref{fig:LdivB}d) is straightforward using the methods we have discussed above, but 
additional subtleties arise for \ref{fig:Qdiv}a) and \ref{fig:Qdiv}b).

When restricting to Higgs fluctuations in the loop, these methods
require analysis of a single second-order differential equation for
the factor function of the Jost solution,
\begin{equation}
\frac{1}{\rho}\partial_\rho \rho\partial_\rho\overline{\eta}_\ell=
2tZ_\ell(t\rho)\partial_\rho \overline{\eta}_\ell
+\frac{1}{\rho^2}\left[g^2(\rho)-2\ell g(\rho)\right]\overline{\eta}_\ell
+V_H(\rho)\overline{\eta}_\ell
\label{eq:jost2}\end{equation}
with
$Z_\ell(z)=\frac{K_{|\ell|+1}(z)}{K_{|\ell|}(z)}-\frac{|\ell|}{z}$ and 
$V_H(\rho)=3(h^2(\rho)-1)$. Let $\overline{\eta}^{(1,2)}_\ell$ be the
solutions for $\overline{\eta}_\ell$ at 
first- and second-order in $V_H(\rho)$, but zeroth order in the 
gauge field. Thus  $\overline{\eta}^{(1,2)}_\ell$ relate to the diagrams \ref{fig:Qdiv}c) 
and \ref{fig:LdivB}d). The scattering data analog of diagram \ref{fig:Qdiv}a) is governed 
by the differential equation
\begin{equation}
\frac{1}{\rho}\partial_\rho \rho\partial_\rho\overline{\eta}^{(3)}_\ell
=2tZ_\ell(t\rho)\partial_\rho \overline{\eta}^{(3)}_\ell
+\left(\frac{g(\rho)}{\rho}\right)^2\,,
\label{eq:BornH3}\end{equation}
while diagram \ref{fig:Qdiv}b) is associated with a set of coupled differential equations
\begin{align}
\frac{1}{\rho}\partial_\rho \rho\partial_\rho\overline{\eta}^{(4)}_\ell
&= 2tZ_\ell(t\rho)\partial_\rho \overline{\eta}^{(4)}_\ell
-\frac{2\ell}{\rho^2}g(\rho)\,,\cr
\frac{1}{\rho}\partial_\rho \rho\partial_\rho\overline{\eta}^{(5)}_\ell
&= 2tZ_\ell(t\rho)\partial_\rho \overline{\eta}^{(5)}_\ell
-\frac{2\ell}{\rho^2}g(\rho)\overline{\eta}^{(4)}_\ell\,.
\label{eq:BornH4}\end{align}
The boundary conditions are such that $\overline{\eta}_\ell\to1$ while all 
$\overline{\eta}^{(i)}_\ell\to0$ as $\rho\to\infty$. Numerical simulations \cite{Graham:2019fzo} 
for regular profiles, {\it i.e.} Eq.~(\ref{eq:bcANO}) replaced by $g(\rho)\sim0$ at 
the center of the vortex, verify that for sufficiently large $t$
\begin{equation}
\sum_{\ell=-\infty}^\infty\lim_{\rho\to0}\left\{
{\rm ln}\left(\overline{\eta}_\ell\right)
-\overline{\eta}^{(1)}_\ell-\overline{\eta}^{(2)}_\ell
+\fract{1}{2}\left(\overline{\eta}^{(1)}_\ell\right)^2
\hspace{-1mm}-\overline{\eta}^{(3)}_\ell-\overline{\eta}^{(4)}_\ell
-\overline{\eta}^{(5)}_\ell
+\fract{1}{2}\left(\overline{\eta}^{(4)}_\ell\right)^2\right\}\,\propto\,\frac{1}{t^4}\,.
\label{eq:BornH5}
\end{equation}
This guarantees a finite integral in Eq.~(\ref{eq:inter2}). More surprisingly,
those numerical experiments also verify that
\begin{equation}
\sum_{\ell=-\infty}^\infty\lim_{\rho\to0}\left\{
\overline{\eta}^{(3)}_\ell+\overline{\eta}^{(4)}_\ell
\hspace{-1mm}+\overline{\eta}^{(5)}_\ell
-\fract{1}{2}\left(\overline{\eta}^{(4)}_\ell\right)^2\right\}
\,\longrightarrow\, n^2\int_0^\infty \frac{d\rho}{\rho}\, g^2(\rho)
\quad {\rm as}\quad t\,\to\,\infty\,,
\label{eq:LinInt}\end{equation}
signaling a quadratic divergence. It is exactly the quadratic divergence that
would emerge from diagrams \ref{fig:Qdiv}a) and b) if the loop were not regularized
in a gauge-invariant manner. A gauge invariant treatment, however, should lead to 
only a logarithmic divergence, reflected by an asymptotic $\frac{1}{t^2}$ behavior.
We recall that Eq.~(\ref{eq:inter2}) originated from integrating 
$\frac{d\left[\nu(t)\right]_N}{dt}$ by parts after the analytic continuation of 
Eq.~(\ref{eq:inter1}). Hence the subtraction of that integral from $\left[\nu(t)\right]$
does not alter the VPE, but rather it restores gauge invariance of the expression in 
Eq.~(\ref{eq:inter2}). 

Unfortunately, the integral in Eq.~(\ref{eq:LinInt}) does not exist for the singular vortex profile. 
Similarly the (first order) Born approximation does not exist. However, we may consider
\begin{equation}
\left[\nu(t)\right]_{H}=\lim_{L\to\infty}\sum_{\ell=-L}^{L}
\left\{{\rm ln}\left(\overline{\eta}_\ell\right)
-\overline{\eta}^{(1)}_\ell-\overline{\eta}^{(2)}_\ell
+\fract{1}{2}\left(\overline{\eta}^{(1)}_\ell\right)^2\right\}
\Bigg|_{\rho=\rho_{\rm min}}
-n^2\int_{\rho_{\rm min}}^\infty \frac{d\rho}{\rho}\, g^2(\rho) \,,
\label{eq:sing1}\end{equation}
which subtracts the gauge invariant logarithmic divergence of
diagrams \ref{fig:Qdiv}a) and b) when $\rho_{\rm min}\to0$, even for
singular backgrounds.  Here the $H$ subscript denotes the $V_H$
subtraction at linear and quadratic orders.  The numerical simulations
confirm that indeed $\left[\nu(t)\right]_{H}$ does not diverge in
that limit.

Using dimensional regularization ($D=4-2\epsilon)$ the logarithmic
divergence in the combination of diagrams \ref{fig:Qdiv}a) and b) is
\begin{equation}
E^{(A)}\Big|_{\rm div.}=
\frac{1}{12\epsilon(4\pi)^2}\int d^2x\, F_{\mu\nu}F^{\mu\nu}
=\frac{1}{96\pi^2}\left[\int d^2x\, F_{\mu\nu}F^{\mu\nu}\right]
\int_0 \frac{l^2 dl}{\sqrt{l^2+M^2}^3}\Bigg|_{\rm div.}\,.
\label{eq:FDA1}\end{equation}
Hence we expect that 
\begin{equation}
\lim_{\rho_{\rm min}\to0} \left[\nu(t)\right]_{H}
\,\longrightarrow\,\nu_{\rm l.f.}(t)=
\frac{1}{t^2}\,\frac{n^2}{12}\int_0^\infty \rho\,d\rho
\left(\frac{g^\prime(\rho)}{\rho}\right)^2
\quad {\rm as}\quad t\,\rightarrow\,\infty\,.
\label{eq:lf3}\end{equation}
To simplify the simulation we employ a one-parameter ($\alpha$) set of 
trial profile functions 
\begin{equation}
h(\rho)=\tanh(\alpha\rho)
\qquad{\rm and}\qquad
g(\rho)={\rm e}^{-(\alpha\rho)^2}
\label{eq:profile}\end{equation}
that reflect the singular structure of the vortex appropriately. Numerically we cannot take 
the angular momentum sum in Eq.~(\ref{eq:sing1}) to infinity. Rather we consider $\nu_L(t)$ as 
the $\rho_{\rm min}\to0$ limit of the right-hand side evaluated with
finite limits
($-L,\ldots,+L$) on the sum.
\begin{figure}
\centerline{
\includegraphics[width=3.8cm,height=3.8cm]{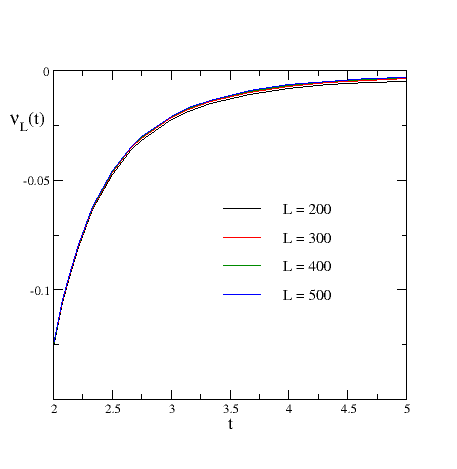}\hspace{0.5cm}
\includegraphics[width=3.8cm,height=3.8cm]{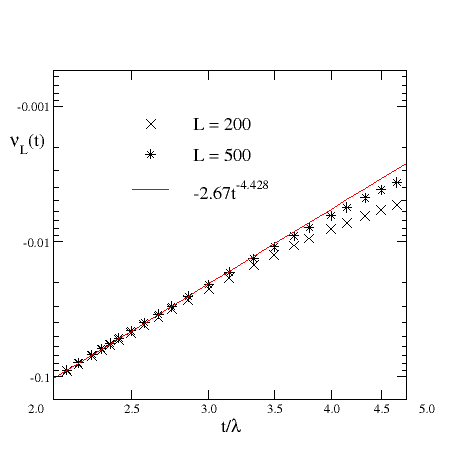}\hspace{0.5cm}
\includegraphics[width=3.8cm,height=3.8cm]{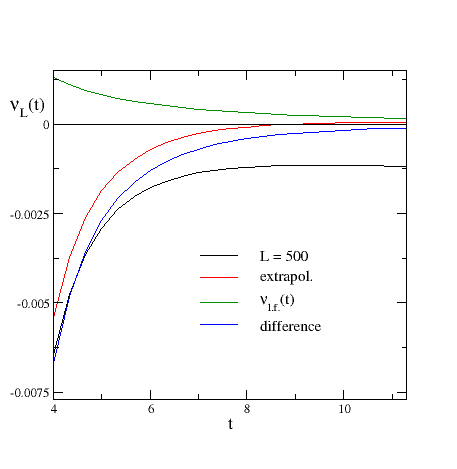}}
\caption{Asymptotic behavior of $\nu_L(t)$ for $\alpha=0.5$ 
in Eq.~(\ref{eq:profile}) and $n=1$. Left panel: different maximal
angular momenta $L$;  center panel: fit showing that $\nu_L(t)$ falls
faster than $1/t^2$ for moderate values of $t$; right panel: extrapolation
$L\to\infty$ and limiting function from Eq.~(\ref{eq:lf3}).}
\label{fig:num1}
\end{figure}
In order to reach the asymptotic behavior, these limits must increase with the (imaginary) 
momentum $t$. In the left panel of Fig.~\ref{fig:num1} we display $\nu_L(t)$ for 
values as large as $L=500$. This graph indeed suggests convergence of the 
angular momentum sum for moderate values of $t$. However, there are at least two problems.
First, the asymptotic value seems to be negative, while $\nu_{\rm l.f.}(t)>0$. Second, 
$\nu_L(t)$ approaches zero approximately like $\frac{1}{t^4}$, as the middle panel of 
Fig.~\ref{fig:num1} suggests. If correct, it would imply that the integral 
$\int_{\sqrt{2}}^\infty dt\,\nu_L(t)$ is finite\footnote{This has caused confusion 
in earlier publications, {\it cf.} Refs. \cite{Pasipoularides:2000gg,Graham:2004jb}.} and 
that the counterterm for the logarithmic divergence from the diagrams~\ref{fig:Qdiv}a) 
and~b) would not be compensated. It turns out that for momenta as small as $t\approx6$,
$\nu_{500}(t)$ has not reached the asymptotic value, as clearly seen in the right panel 
of Fig.~\ref{fig:num1}. Numerically $L>600$ is difficult to handle and costly in 
CPU-time because of the singular behavior of the modified Bessel functions at small 
arguments. Instead an extrapolation from $L\in[300,600]$ to infinity
is required. Indeed, 
that infinite $L$ extrapolation turns positive at large $t$ and the difference from 
$\nu_{\rm l.f.}(t)$ is numerically confirmed to decay faster than $\frac{1}{t^3}$.
The results shown in Fig.~\ref{fig:num1} are for topological charge
$n=1$, but the cases $n=2,3,4$ have been confirmed to follow the same behavior.

We have thus shown that both the subtraction of the constant in Eq.~(\ref{eq:lf3}) and
the extrapolation to infinite angular momentum are necessary to comply
with gauge invariance.

\subsection{VPE for different topological charges}
\label{ssec:topcharge}

\begin{figure}[t]
\centerline{
\includegraphics[width=11cm,height=2.5cm]{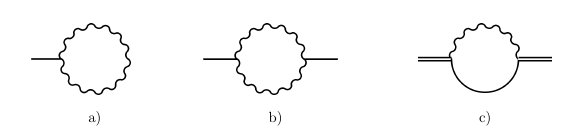}}
\caption{Additional divergent one-loop diagrams in the presence of gauge fluctuations.
The double line indicates the insertion of the off-diagonal
interaction $d(\rho)$.}
\label{fig:Gdiv}
\end{figure}

We want to adopt the procedure established above to the VPE of the ANO vortex, with 
$\nu_\ell(t)=\lim_{\rho\to0}{\rm ln} \,{\rm det}\left[\mathcal{F}_\ell\right]$ where
$\mathcal{F}_\ell$ is matrix solution to Eq.~(\ref{eq:jostdeq}). Since we must 
separate the singular terms of the Higgs-Higgs component of the
potential matrix, we introduce
\begin{equation}
\overline{\mathcal{V}}=\begin{pmatrix}
3(h^2(\rho)-1) & \sqrt{2}d(\rho)\cr
\sqrt{2}d(\rho)& 2(h^2(\rho)-1) \end{pmatrix}
\label{eq:potmat1}\end{equation}
and let $\overline{\nu}_\ell^{(1)}$ and $\overline{\nu}_\ell^{(2)}$ the first two Born terms 
originating from $\overline{\mathcal{V}}$. They are the scattering data analog of the
diagrams \ref{fig:Qdiv}c), \ref{fig:LdivB}d) and those in Fig.~\ref{fig:Gdiv}. The above 
analysis suggests that\footnote{In the numerical analysis an extrapolation as in 
Eq.~(\ref{nuextra}) is needed for channels with $|\ell|\le n$. This minor ($\sim1\%$) correction
was not included in Ref.~\cite{Graham:2021mbc}.}
\begin{equation}
\left[\nu(t)\right]_V=
\lim_{\genfrac{}{}{0pt}{}{L\to\infty}{\rho_{\rm min}\to0}}\left\{
\sum_{\ell=-L}^{L}\left[\nu_\ell(t)-\overline{\nu}^{(1)}_\ell(t)
-\overline{\nu}^{(2)}_\ell(t)\right]_{\rho_{\rm min}}
-n^2\int_{\rho_{\rm min}}^\infty \frac{d\rho}{\rho}\,g^2(\rho)\right\}
\label{eq:sing2}
\end{equation}
approaches~~$\displaystyle \frac{n^2}{12t^2}\int_0^\infty \rho\,d\rho
\left(\frac{g^\prime(\rho)}{\rho}\right)^2$ as $t\to \infty$. We treat the resulting 
logarithmic divergence using the fake boson method introduced in Sec.~\ref{ssec:fake}.
Then the scattering part of the VPE reads
\begin{equation}
\Delta E_{\rm scat.}=\frac{1}{2\pi}\int_{\sqrt{2}}^\infty tdt\,
\left[\left[\nu(t)\right]_V-C_B \nu_B^{(2)}(t)\right]\,,
\label{eq:Escat}
\end{equation}
and for the choice $V_B(\rho)=3(\tanh^2(\zeta \rho)-1)$ the fake boson
coefficient becomes
\begin{equation}
C_B=-\frac{n^2}{3}\frac{\int_0^\infty \rho d\rho
\left(\frac{g^\prime(\rho)}{\rho}\right)^2}
{\int_0^\infty \rho d\rho\, \left[3(\tanh^2(\zeta \rho)-1)\right]^2}\,.
\label{eq:CB}\end{equation}
Here $\zeta$ is a tunable parameter that has no effect once the renormalized Feynman diagram, 
which is obtained from two insertions of $V_B$ and subsequent multiplication by $C_B$, is added. 

The counterterm Lagrangian reads
\begin{equation}
\mathcal{L}_{\rm CT}=c_1F_{\mu\nu}F^{\mu\nu}+c_2|D_\mu\Phi|^2
-c_3\left(|\Phi|^2-v^2\right)^2-c_4\left(|\Phi|^2-v^2\right)\,.
\label{eq:lct}\end{equation}
The first three terms are from the original Lagrangian, Eq.~(\ref{eq:lag1}). 
The last term cancels the tadpole diagrams and is generated from the original Lagrangian
by varying the vacuum expectation value $v$. Hence the no-tadpole condition fixes $c_4$
and ensures that $v$ has no quantum corrections. As in Sec.~\ref{ssec:onshell}, we impose 
on-shell conditions to fix the counterterm coefficients $c_1$, $c_2$, and $c_3$.  This requires
consideration of the diagrams \ref{fig:Qdiv}a), \ref{fig:Qdiv}b), \ref{fig:LdivB}d) as well as 
\ref{fig:Gdiv}b) and \ref{fig:Gdiv}c). For example, in momentum space the diagrams \ref{fig:Qdiv}a) 
and b) yield the dimensionally regularized action (returning to physical units
with $M=\sqrt{2}ev$)
$$
\Gamma\left(1-\fract{D}{2}\right)
\left(\fract{eM}{4\pi}\right)^2\left(\fract{4\pi\mu^2}{M^2}\right)^{2-\frac{D}{2}}
\int \frac{d^4p}{(2\pi)^4}\,\widetilde{A}_\mu(p)\widetilde{A}^\mu(-p)
\int_0^1dx\left[1-X(x,p)^{\frac{D}{2}-1}\right]\,,
$$
where $X(x,p)=1-x(1-x)\fract{p^2}{M^2}$.
By gauge invariance there is no pole at $D=2$ and we can analytically continue
to $D=4-2\epsilon$ with the renormalization scale $\mu^2$
\begin{align*}
&-\fract{e^2}{96\pi^2}\left[\frac{1}{\epsilon}+1-\gamma
+{\rm ln}\fract{4\pi\mu^2}{M^2}\right]
\int \frac{d^4p}{(2\pi)^4}\,p^2\widetilde{A}_\mu(p)\widetilde{A}^\mu(-p)
\cr
&\hspace{1.2cm}
+\left(\fract{eM}{4\pi}\right)^2\int\frac{d^4p}{(2\pi)^4}\widetilde{A}_\mu(p)\widetilde{A}^\mu(-p)
\int_0^1dx\,X(x,p) \,{\rm ln}\left[X(x,p)\right]\,.
\end{align*}
Combining this result with the first term in Eq.~(\ref{eq:lct}) and
separating the  divergent part of the counterterm coefficient via
$2c_1=\fract{e^2}{96\pi^2}\left[\frac{1}{\epsilon}+1-\gamma
+{\rm ln}\fract{4\pi\mu^2}{M^2}\right]+c_A$ yields\footnote{Note that for 
the vortex $\partial_\mu A^\mu_S=0$ and thus $p_\mu\widetilde{A}^\mu(p)=0$. Also note 
that ${\rm ln}\left[X(x,0)\right]=0$ so that $G_A(p^2)$ is regular when $p^2\to0$.}
$$
\int \frac{d^4p}{(2\pi)^4}\,p^2\widetilde{A}_\mu(p)\widetilde{A}^\mu(-p)G_A(p^2)
$$
where
$$
G_A(p^2)=c_A+\left(\frac{eM}{4\pi}\right)^2\frac{1}{p^2}\int_0^1dx\,X(x,p)
\,{\rm ln}\left[X(x,p)\right]\,.
$$
Quantum corrections are eliminated from the residue of the gauge field propagator
by setting $G_A(M^2)=0$, which determines $c_A$.
The condition on the residue of the Higgs propagator determines $c_2$ from 
diagram \ref{fig:Gdiv}c), while the combination of the diagrams \ref{fig:Qdiv}c)
and \ref{fig:Gdiv}b) yields $c_3$ by demanding that the Higgs
mass is not changed by quantum effects. It is worth noting that the equality of Higgs 
and gauge field masses is maintained when one-loop quantum corrections are included.

With all counterterm coefficients determined, we can write the (renormalized) 
Feynman diagram piece of the VPE as
\begin{equation}
E^{\rm ren.}_{\rm FD}=C_B E_{\rm FD}^{(2)} + E_{\rm FD}^{(0)}+E_{\rm CT}\,.
\label{eq:Efd}
\end{equation}
The Feynman diagram with two insertions of $V_B$ gives $E_{\rm FD}^{(2)}$, while 
$E_{\rm FD}^{(0)}$ originates from diagrams \ref{fig:Qdiv}c), \ref{fig:LdivB}d), 
\ref{fig:Gdiv}b) and \ref{fig:Gdiv}c). By construction, $C_B E_{\rm FD}^{(2)}$ 
has the same ultra-violet divergence as the combination of the diagrams \ref{fig:Qdiv}a) 
and~b). Hence all such divergences cancel in $E^{\rm ren.}_{\rm FD}$.

We list the numerical results in Tab.~\ref{tab:VPE_ANO}. We have also performed
simulations with different values for $\zeta$ and ensured equal results for $\Delta E$ to 
the given precision. (Of course, the individual contributions $E^{\rm ren.}_{\rm FD}$ and
$\Delta E_{\rm scat.}$ vary with $\zeta$.)
\begin{table}[ht]
\tbl{\label{tab:VPE_ANO}VPE of ANO vortices with different topological charges. The 
parameters in the fake boson potential are $\zeta=1,0.9,0.8,0.7$ for $n=1,2,3,4$, respectively.}
{\begin{tabular}{c||c|c|c|c}
& $n=1$ & $n=2$ & $n=3$ & $n=4$ \cr
\hline
$\Delta E_{\rm scat.}$     &~-0.0510~&~-0.1937~&~-0.3563~&~-0.5251 \cr
$E^{\rm ren.}_{\rm FD}$    &~0.0448~&~0.0558~&~0.0840~&~0.1171 \cr
\hline
$\Delta E $                &~-0.0063~&~-0.1379~&~-0.2722~&~-0.4080 \cr
\end{tabular}}
\end{table}
We observe a constant increment of $\Delta E$ with the topological charge. More precisely,
the fit $\Delta E(n)=-0.005-0.134(n-1)$ has very small $\chi^2=4.5\times10^{-6}$. Since the 
classical energy is linear in $n$ for the case of equal Higgs and gauge field masses, the total 
binding energy, $E_{\rm rel}=\Delta E(n)-n\Delta E(1)=-0.129(n-1)$; is negative. This 
suggests that these vortices coalesce.

Though this is the first computation of the VPE for soliton-like configurations with different
topological charges in a renormalizable theory, it is merely the beginning for computing 
the VPE of vortices. The case with two space dimensions will be next
step. Subsequently the case of different
Higgs and gauge field masses should also be considered.

\section{Summary}

In this short review we have reported recent progress on the computation of 
vacuum polarization energies (VPE) of soliton-like structures in renormalizable quantum field 
theories using spectral methods. For earlier applications of these methods we refer to
the lecture notes of Ref.~\cite{Graham:2009zz}.

For static background configurations like solitons, quantum fluctuations obey wave equations analogous 
to those in ordinary quantum mechanics, with a potential induced by the soliton. Spectral methods then 
determine the bound state energies and scattering data ($S$-matrix, phase shifts) for that potential. 
Formally these data yield the VPE as a sum of the bound state energies and a momentum integral over 
the phase shift, but it is a subtle problem to unambiguously combine the ultra-violet divergences with 
the counterterms of the quantum theory. These counterterm coefficients are universal for 
a prescribed set of renormalization conditions and not sensitive to the particular soliton 
configuration. This problem is solved by the observation that there
are two equivalent expansions for the VPE in powers of the background potential: (i) the sum of Feynman diagrams in 
the quantum field theory and (ii) the Born series for scattering data. In both cases, the series expansion 
approaches the ultra-violet behavior of the full result. To implement the renormalization procedure, the
relevant Born terms are then subtracted from the integrand of the momentum integral and added back in form 
of the equivalent Feynman diagrams, which are then unambiguously combined with the counterterms using
standard techniques. 

Here we have focused on the use of the analytic properties of the scattering data, represented by 
the Jost function, to evaluate the momentum integral along the imaginary momentum axis. This approach has,
among others, the advantage that the bound state energies need not be explicitly computed. We 
have shown that for popular soliton models in one dimension this
approach allows one to compute the VPE very 
efficiently. Moreover, with an appropriate treatment of the mass gap when analytically 
continuing, the VPE for soliton models with quantum fluctuations with different masses can be 
evaluated as well. We stress that the Jost function is not computed in any approximation and that 
its Born expansion solely serves to identify the ultra-violet divergences from the quantum field
theory in the scattering data.
\bigskip

These enhanced spectral methods have made possible the computation of the VPE of more intricate 
configurations in particular quantum field theories. In turn, these studies lead to a number of 
interesting and novel observations: 
\begin{itemize}
\itemsep2mm
\item[$\bullet$]
In models with one space dimension we have documented the novel effect of quantum destabilization of 
solitons. The conjecture is that in theories with multiple (distinct) vacua in field space that have 
different curvatures of the field potential, the soliton may approximately assume either of these vacuum 
configurations in separated regions. As the sizes of these regions
vary, the VPE, and, as a result, the total energy, may decrease without 
a lower bound. We have demonstrated this scenario explicitly 
for the Shifman-Voloshin soliton, which has two field components with different masses. Models with 
field potentials of higher polynomial order also support this conjecture, though there the situation is 
less stringent because the two separated regions both extend to spatial infinity (positive and negative).
\item[$\bullet$]
We have also expressed the interface formalism, which was first developed to describe domain walls
or surfaces subject to the Casimir force, in terms of imaginary momentum integrals. This approach
made possible the computation of the energy carried by fermions in the background of cosmic
strings. At the center of the string, the gauge field has singularities. Fortunately, for 
non-abelian strings the fields are single-valued and the singularity can be removed by a 
local gauge transformation. Neither the components of the Born series nor individual Feynman
diagrams are gauge invariant, and hence the individual ingredients of the spectral method are not 
manifestly gauge invariant. We have therefore made considerable numerical efforts to
confirm gauge invariance of the final results and thereby corroborate the enhanced spectral 
methods. These results show that the total fermion contribution to the energy may be negative 
for certain string profiles, but with magnitude that is too small to overcome the
classical bosonic energy and bind the cosmic string in an $SU_L(2)$
gauge theory. However, when taking the fermion mass as about twice the
top quark mass (or larger), the population of bound state levels yields
a  total energy that is less than the energy of equally many free
fermions. Such a configuration is thus bound and can be viewed as a
new solution in a model similar to the Standard Model 
of particle physics.
\item[$\bullet$]
The enhanced spectral methods have furthermore enabled the first computations of the 
VPE of solitons with different topological charges in a renormalizable quantum field theory,
for Abrikosov-Nielsen-Olesen vortices in the BPS version of the Abelian Higgs model with
spontaneous symmetry breaking in four space-time dimensions. Here additional subtleties arise 
because the singular structure of the vortex cannot be removed and hampers the construction 
of the Born series. An alternative procedure to extract the ultra-violent divergences is needed,
in particular, for the quadratic divergence that emerges from gauge-variant components 
that eventually cancel out in the final result. The correct (subleading) logarithmic divergence was 
obtained by subtracting a momentum-independent constant whose net contribution is zero but renders 
consistency with gauge invariance. Technically that constant is tricky to identify because it is 
only defined in the limit as the wave equations are solved in close vicinity of the vortex. 
Even after proper subtraction of that constant, the singular structure causes the angular momentum sum 
to converge only very slowly and an additional extrapolation is needed in numerical simulations. 
In the BPS case of equal gauge and scalar masses, these calculations show that the VPE is essentially 
proportional to a constant plus a linear function of the topological charge with negative slope.
Since the classical energy is strictly linear in the charge, the vortex  with a given charge has less 
energy than equally many vortices of unit charge and is thus stable, leading to type I behavior of
superconductors in the BPS case.
\end{itemize}

Several possible extensions of these results are of interest. The restriction to the BPS case of equal 
masses reduces the wave equations to a two-channel problem, while in general there will be four coupled 
channels. Also the  case with two space dimensions differs from the calculation in three space dimensions 
because it does not have a complete cancellation between the contributions from the ghost fields (needed for 
gauge fixing) and the un-physical gauge field components. It is thus a five channel scattering problem. In 
addition, the divergence structure is modified by the lower dimensionality.  Though these generalizations
are tractable, their implementation is a worthwhile future project. 

\section*{Acknowledgments}

N.\@ G.\@ is supported in part by the National Science Foundation (NSF)
through grant PHY-1820700.
H.\@ W.\@ is supported in part by the National Research Foundation of
South Africa (NRF) by grant~109497.

\bibliographystyle{ws-ijmpa}

\end{document}